\documentclass{jfm}
\pdfoutput=1 
\usepackage{graphicx}
\usepackage{natbib}
\usepackage{subfig}
\usepackage[amssymb]{SIunits}
\usepackage{amsmath}
\usepackage{color}
\usepackage{epstopdf}

\ifCUPmtlplainloaded \else
  \checkfont{eurm10}
  \iffontfound
    \IfFileExists{upmath.sty}
      {\typeout{^^JFound AMS Euler Roman fonts on the system,
                   using the 'upmath' package.^^J}%
       \usepackage{upmath}}
      {\typeout{^^JFound AMS Euler Roman fonts on the system, but you
                   dont seem to have the}%
       \typeout{'upmath' package installed. JFM.cls can take advantage
                 of these fonts,^^Jif you use 'upmath' package.^^J}%
      }
  \else
  \fi
\fi

\ifCUPmtlplainloaded \else
  \checkfont{msam10}
  \iffontfound
    \IfFileExists{amssymb.sty}
      {\typeout{^^JFound AMS Symbol fonts on the system, using the
                'amssymb' package.^^J}%
       \usepackage{amssymb}%
         \let\leq=\leqslant
         \let\geq=\geqslant
      }{}
  \fi
\fi

\ifCUPmtlplainloaded \else
  \IfFileExists{amsbsy.sty}
    {\typeout{^^JFound the 'amsbsy' package on the system, using it.^^J}%
     \usepackage{amsbsy}}
    {\providecommand\boldsymbol[1]{\mbox{\boldmath $##1$}}}
\fi

\newcommand{\nom}{Nu_\omega}

\newcommand{\usro}{Ro^{-1}}
\newcommand{\bu}{\boldsymbol{u}}

\newsavebox{\astrutbox}
\sbox{\astrutbox}{\rule[-5pt]{0pt}{20pt}}

\title[Exploring the phase diagram of fully turbulent Taylor-Couette flow]{Exploring the phase diagram of fully turbulent Taylor-Couette flow}

\author[R. Ostilla M\'onico and others]%
{R\ls O\ls D\ls O\ls L\ls F\ls O\ns O\ls S\ls T\ls I\ls L\ls L\ls A\ns M\ls \'O\ls N\ls I\ls C\ls O$^1$,\break%
E\ls R\ls W\ls I\ls N\ns P.\ns V\ls A\ls N\ns D\ls E\ls R\ns P\ls O\ls E\ls L$^1$,\ns%
R\ls O\ls B\ls E\ls R\ls T\ls O\ns V\ls E\ls R\ls Z\ls I\ls C\ls C\ls O$^{2,1}$,\break%
S\ls I\ls E\ls G\ls F\ls R\ls I\ls E\ls D\ns G\ls R\ls O\ls S\ls S\ls M\ls A\ls N\ls N$^{3}$,\ns%
\and D\ls E\ls T\ls L\ls E\ls F\ns L\ls O\ls H\ls S\ls E$^1$}

\affiliation{$^1$Physics of Fluids, Mesa+ Institute, University of Twente, P.O. Box 217, 7500 AE Enschede, The Netherlands\\[\affilskip]
$^2$Dipartimento di Ingegneria Meccanica, University of Rome ``Tor Vergata'', Via del Politecnico 1, Roma 00133, Italy\\[\affilskip]
$^3$Department of Physics, University of Marburg, Renthof 6, D-35032 Marburg, Germany}

\date{\today}

\begin{document}
 
\maketitle

\begin{abstract}
Direct numerical simulations of Taylor-Couette flow (TC), i.e. the flow between two coaxial
and independently rotating cylinders were performed. Shear Reynolds numbers of
up to $3\cdot10^5$, corresponding to Taylor numbers of $Ta=4.6\cdot10^{10}$, were
reached. Effective scaling laws for the torque are presented. 
The transition to the ultimate regime, in which asymptotic scaling laws (with logarithmic corrections) for the
torque are expected to hold up to arbitrarily high driving, is analysed for different radius ratios,
different aspect ratios and different rotation
ratios. It is shown that the transition is approximately independent of the aspect- and rotation- ratios,
but depends significantly on the radius-ratio. We furthermore calculate the
local angular velocity profiles and visualize different flow regimes
that depend both on the shearing of the flow,
and the Coriolis force originating from the outer cylinder rotation. Two main regimes
are distinguished, based on the magnitude of the Coriolis force, namely the co-rotating
and weakly counter-rotating regime dominated by Rayleigh-unstable regions, and the strongly
counter-rotating regime where a mixture of Rayleigh-stable and Rayleigh-unstable regions exist.
Furthermore, an analogy between radius-ratio and outer-cylinder rotation is revealed, namely that
smaller gaps behave like a wider gap with co-rotating cylinders, and that wider gaps behave like
smaller gaps with weakly counter-rotating cylinders.
Finally, the effect of the aspect ratio on the effective torque versus Taylor number
scaling is analysed and it is shown that different branches of the torque-versus-Taylor
relationships associated to different aspect ratios are found to cross within $15\%$ of the Reynolds number associated to the transition to the ultimate regime.
The paper culminates in phase diagram in the inner vs outer Reynolds number parameter space and in the Taylor
vs inverse Rossby number parameter space, which can be seen as the extension of the Andereck \emph{et al.}
(J. Fluid Mech. 164, 155-183, 1986) phase diagram towards the ultimate regime.
\end{abstract}


\section{Introduction}

Taylor-Couette flow (TC), i.e. the flow between two independently rotating concentric cylinders, has 
for long been used as a model system in fluid dynamics. \cite{cou890} was the first to investigate it,
and he pioneered its usage as a viscometer. But it was \cite{mal896} who, by rotating the inner cylinder,
and not the outer as Couette had done, found the first indications of turbulence in the system. 
\cite{tay23,tay36b} further studied the system, finding that it was linearly unstable, unlike
pipe-flow and other studied systems to the date. \cite{wen33}
expanded the study of the turbulent regime, measuring torques and velocities in the system. 
Since then, and due to its simplicity, TC has been used as a model system for studying shear flows. 
For a broader historical context, we refer the reader to \cite{don91b}.

Recently, a mathematically exact analogy between TC and Rayleigh-B\'enard flow (RB), i.e.
the convective flow between two parallel plates heated from below and cooled from above was found by Eckhardt,
Grossmann and Lohse (2007), (here referred to as EGL07) \nocite{eck07b}. 
Within this context, TC can be viewed as a convective flow, driven by the shear between both
cylinders where angular \emph{velocity} is transported from the inner 
to the outer cylinder. As explained by \cite{gro14}, as long as the driving of the system is small, the 
transport is limited by the laminar boundary layers. 
But if the driving becomes strong enough the boundary layers become turbulent and the system
enters the so-called ``ultimate'' regime. The study of the transition to this regime, expected to be also present in RB,
has attracted recent interest, as most applications of TC and RB in geo- and astro-physics are expected to be in
this ultimate regime.

For RB flow, the transition to an ultimate regime was first qualitatively 
predicted by \cite{kra62}, and later
quantitatively by \cite{gro00,gro01,gro11} and then experimentally 
found by \cite{he12,he12a,ahl12b,roc10}. 
It lies outside the present reach of DNS. 
The analogous boundary layer transition to an ultimate regime in TC flow was first found in the experiments by \cite{lat92,lat92a}, and analysed more precisely in \cite{lew99}, 
even though earlier work by \cite{wen33} already showed some transition
in the torque scaling around the same Reynolds number. 
The transition was not related to the transition to the ultimate regime until later \citep{gil11,pao11,hui12,gro14}. In DNS, it was observed for the first time in \citep{ost14}.
 
In TC flow this transition is easier to achieve as the mechanical driving is more efficient than the thermal one,
and thus the frictional Reynolds numbers in the boundary layer are much larger.
By using the analogy between both systems, better understanding of the transition in TC can thus also lead 
to new insight in RB, where it is more elusive.

\cite{ost14} numerically studied the transitions in TC for pure inner cylinder rotation for a radius ratio of $\eta=r_i/r_o=0.714$, 
where $r_o$ and $r_i$ are the outer and inner radii respectively, and an aspect ratio $\Gamma=L/(r_o-r_i)=2\pi/3$, where $L$ is the
axial period in the DNS.
In that study, the flow transitions and boundary layer dynamics were revealed in the range of Taylor numbers $Ta$ between $10^4$ and $10^{10})$, where the Taylor number is defined as:

\begin{equation}
 Ta=\frac{1}{4}\sigma d^2(r_o+r_i)^2(\omega_i-\omega_o)^2\nu^{-2}, 
\end{equation}

\noindent with $\omega_o$ and $\omega_i$ the angular velocities 
of the outer and inner cylinder, respectively, $d=r_o-r_i$ the gap width, and $\nu$ the kinematic viscosity 
of the fluid. $\sigma=[(r_o+r_i)/(2\sqrt{r_ir_o})]^4$ can be considered as a geometric quasi-Prandtl number (EGL07).

We now describe the series of events when increasing $Ta$. For small enough $Ta$, the flow is 
in the purely azimuthal, laminar, state. When the system is driven beyond a critical driving, one passes the onset of instability and
the purely azimuthal, laminar, flow disappears and large-scale Taylor rolls form. Further increasing of the driving 
breaks up these rolls, causing the onset of time-dependence as the system transitions from the stationary Taylor vortex regime
to the modulated Taylor vortex regime and finally the breakup of these into chaotic turbulent Taylor vortices. 
These changes of the flow are reflected in transitions of the local scaling laws for the torque versus driving, i.e. versus Taylor number $Ta$.
All this has been studied extensively and summarized e.g. in \cite{and83,lat92,lat92a,lew99}. 
The mentioned breakup of the rolls leads to the existence of a transitional regime, where the large-scale coherent 
structures still can be identified when looking at the time-averaged quantities. Looking at the details of the flow, a mixture of turbulent and laminar
boundary layers is present.

In this transitional regime, hairpin vortices, which, in the context of RB, can also be viewed as plumes,
are ejected from both inner and outer cylinders, and these 
contribute to large-scale bulk structures. These structures in turn cause an axial pressure gradient,
which couples back to the boundary layers, causing plumes to be ejected there. But this only happens 
from preferential spots in the boundary layers.
Once the driving is strong enough, the large-scale structures slowly vanish, and the plumes
no longer feel an axial pressure gradient. The boundary layers now 
become fully turbulent and the flow transitions to the ``ultimate'' regime.
As the flow enters the ultimate regime, and the 
boundary layer become turbulent, a logarithmic signature in the \emph{angular} velocity boundary layers is
expected, which indeed has been found experimentally \citep{hui13} and numerically \citep{ost14}.

In the ultimate regime, an effective scaling relation between the Nusselt number $\nom$, i.e.
the non-dimensional torque
$\nom=T/T_{pa}$ where $T$ is the torque, and $T_{pa}$ the torque in the purely azimuthal state, and the Taylor number $Ta$ is expected, with an effective scaling exponent which 
exceeds that for the laminar-type boundary layer case \citep{mal54}, for which $\alpha=1/3$. 
I.e. in the ultimate regime, we expect an effective scaling law $\nom\sim Ta^\alpha$ with $\alpha>1/3$.
In fact, for that regime, the relation law $\nom\sim Ta^{1/2}$ with logarithmic corrections was suggested, \citep{kra62,spi71,gro11}. 
The logarithmic corrections are quite large, and lead to an effective scaling law with $\alpha\approx 0.38$ for $Ta\sim 10^{11}$ \citep{gro11,gil12}. We note that this scaling law is analog to the scaling of the friction factor with
Reynolds number in fully turbulent pipes \cite{pra33}.

For the largest drivings, remnants of the larger rolls, which can be seen as a large scale wind, 
are still observed at even the largest Reynolds numbers studied numerically \citep{ost14}, and experimentally, even up to $Re\sim10^6$ \citep{hui14}.
In \cite{ost14}, the remnants of the large scale structures played a crucial role in the transition to the ultimate regime.
However, large scale structures are not present in the whole parameter space of TC. \cite{and86} showed how rich a variety of 
different states exists at low Reynolds number when the outer cylinder is also rotated.
\cite{bra13b} reported that the strength of the large scale wind was most pronounced at the position of optimal transport. 
However, if the outer cylinder is counter-rotated
past the position of optimal transport, bursts arise from the outer cylinder. The flow is very different
outside and inside the neutral surface, which separates
Rayleigh-stable from Rayleigh-unstable regions of the gap, changing completely the dynamics of the system.
The Taylor vortices no longer penetrate the whole gap, extending thus the unstable region effectively 
somewhat outside the neutral surface of laminar type flow \citep{ost13}.

The geometry of the system can be expected
to play an important role in determining the strength of the large scale wind, and how the transition takes place. In the context of 
understanding the radius-ratio dependence of the transition to the ultimate regime, \cite{mer13} reported a higher
transitional Reynolds numbers for $\eta=0.5$ than 
what was seen for $\eta=0.714$ by \cite{ost14} and for $\eta=0.909$ by \cite{rav10}.
Also the aspect-ratio plays a role. Although different vortical states 
were known to coexist at low Reynolds number \citep{ben78}, it was previously
thought that if the driving was sufficiently large, only one branch of the torque versus Taylor number
curve would survive \citep{lew99}. \cite{bra13}
found that the difference in the global response between different vortical states becomes smaller with increasing Reynolds number.
Recently, \cite{mar14} reported on the existence of different vortical states associated to different global 
torques at a given Taylor number for $\eta=0.909$, 
and that there is a crossing between those torque-versus-$Ta$ curves around the transition to the ultimate regime. Furthermore, \cite{hui14} showed that 
different vortical states survive up to Reynolds number of $10^6$, corresponding to Taylor numbers of order $10^{12}$.
Furthermore, by combining measurements of global torque and local velocity, \cite{hui14}
found that the optimal transport is connected to the \emph{existence} of the large-scale coherent structures at high Taylor numbers. 

Therefore, some questions arise which we want to address in the present paper: How does the transition in the boundary layers
take place across the full parameter space
of TC? Is the vanishing of the large-scale wind a necessary and/or a sufficient condition
for the boundary layer transition? Why does the
transition occur later for $\eta=0.5$ than for larger values of $\eta$?
And finally, what is the effect of the vortical wavelength and why 
do different branches of the torque versus Taylor number scaling curves cross near the transition to the ultimate regime? 

\section{Explored parameter space}

\subsection{Control parameters}

To answer these questions, direct numerical simulations (DNS) of TC have been performed 
across all dimensions of the parameter space, not only adding outer
cylinder rotation, but also varying both geometrical parameters $\eta$ and $\Gamma$.
To do this, the rotating-frame formulation of \cite{ost13} was used. In that paper, TC was formulated 
in a frame rotating with the outer cylinder, such that it looks like a system 
in which only the inner cylinder is rotating, but with a Coriolis force term, which represents the original presence of 
the outer cylinder rotation. The shear driving of the system is non-dimensionally expressed as a Taylor number, introduced previously:

\begin{equation}
 Ta=\frac{1}{4}\sigma d^2(r_o+r_i)^2(\omega_i-\omega_o)^2\nu^{-2}, 
\end{equation}
 
\noindent $Ta$ is the analog to the Rayleigh number in RB, as elaborated in EGL07. The outer cylinder rotation reflects in a Coriolis force, 
characterized by a Rossby number $Ro=|\omega_o-\omega_i|~r_i/(2\omega_od)$.
The Rossby number or rather $Ro^{-1}$ is the parameter which appears in the equations of motion for the fluid:

\begin{equation}
 \frac{\partial \tilde{\bu}}{\partial \tilde{t}} + \tilde{\bu}\cdot\tilde{\nabla}\tilde{\bu} = -\tilde{\nabla} \tilde{p} +  
\displaystyle\frac{f(\eta)}{Ta^{1/2}} \tilde{\nabla}^2\tilde{\bu} - Ro^{-1} {\boldsymbol{e}_z}\times\tilde{\bu}~,
\label{eq:rotatingTC}
\end{equation}

\noindent where $f(\eta)=\frac{1}{4}\sigma((1+\eta)/\eta)^2$, a geometrical parameter. The Rossby number is related to the frequency ratio $\mu = \frac{\omega_o}{\omega_i}$ via 

\begin{equation}
Ro^{-1} = \mbox{sgn} (\omega_o) ~ \left| \frac{\mu}{\mu - 1} \right| ~ \frac{2(1-\eta)}{\eta}~. 
\end{equation} 

\noindent Thus fixed $Ro^{-1}$ means fixed $\mu$ and vice versa. $Ro^{-1} > 0$ describes co-rotation or $\omega_o > 0$, 
while $Ro^{-1}< 0$ means counter-rotation. The radius ratio $\eta$ is presented by the geometrical amplitude factor
$2(1-\eta)/\eta$, being small for small gap ($\eta \rightarrow 1$) and large for large gap ($\eta \rightarrow 0$). A 
resting outer cylinder is described by $Ro^{-1} = 0$. 

There are also other ways of choosing the control parameters. Classically, they have been expressed as two 
non-dimensional Reynolds numbers corresponding to the inner and outer cylinders: $Re_{i,o}=u^\theta_{i,o} \cdot d/\nu$,
where $u^\theta_{i,o}$ are the azimuthal velocities of the inner and outer cylinders. The classical flow control parameters $(Re_i,Re_o)$
can be transformed to the $(Ta,\usro)$ parameter space by:

\begin{equation}
 Ta=f(\eta) |Re_i-\eta Re_o|^2, 
\end{equation}

\noindent and

\begin{equation}
 \usro=\displaystyle\frac{2(1-\eta)Re_o}{|\eta Re_o-Re_i|}.
\end{equation}

\noindent Viceversa, we have

\begin{equation}
 Re_i=\left(\displaystyle\frac{Ta}{f(\eta)}\right)^{1/2}\left ( 1 + \displaystyle\frac{\eta\usro}{2(1-\eta)} \right ),
\end{equation}

\noindent and

\begin{equation}
  Re_o = \displaystyle\frac{\usro Ta^{1/2}}{2f(\eta)^{1/2}(1-\eta)}.
\end{equation}

\noindent The driving can also be expressed as a shear Reynolds number $Re_s=\sqrt{Ta/\sigma}$.

\subsection{Numerical scheme}

\begin{figure}
 \begin{center}
  \subfloat{\label{fig:ReiReoPhasespace}\includegraphics[width=0.49\textwidth,trim = 0mm 0mm 9mm 0mm, clip]{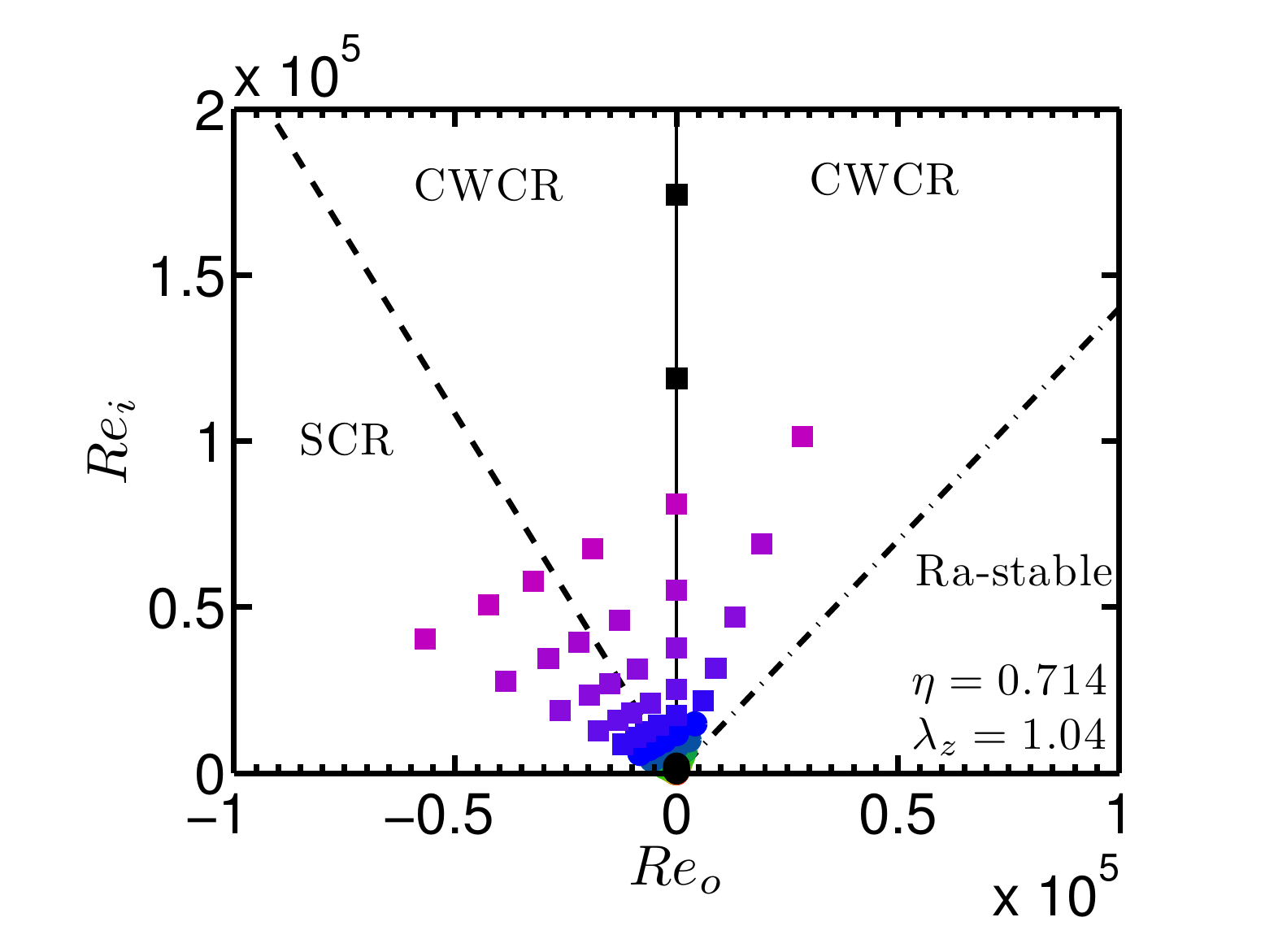}}
  \subfloat{\label{fig:TaRoPhasespace}\includegraphics[width=0.49\textwidth,trim = 0mm 0mm 9mm 0mm, clip]{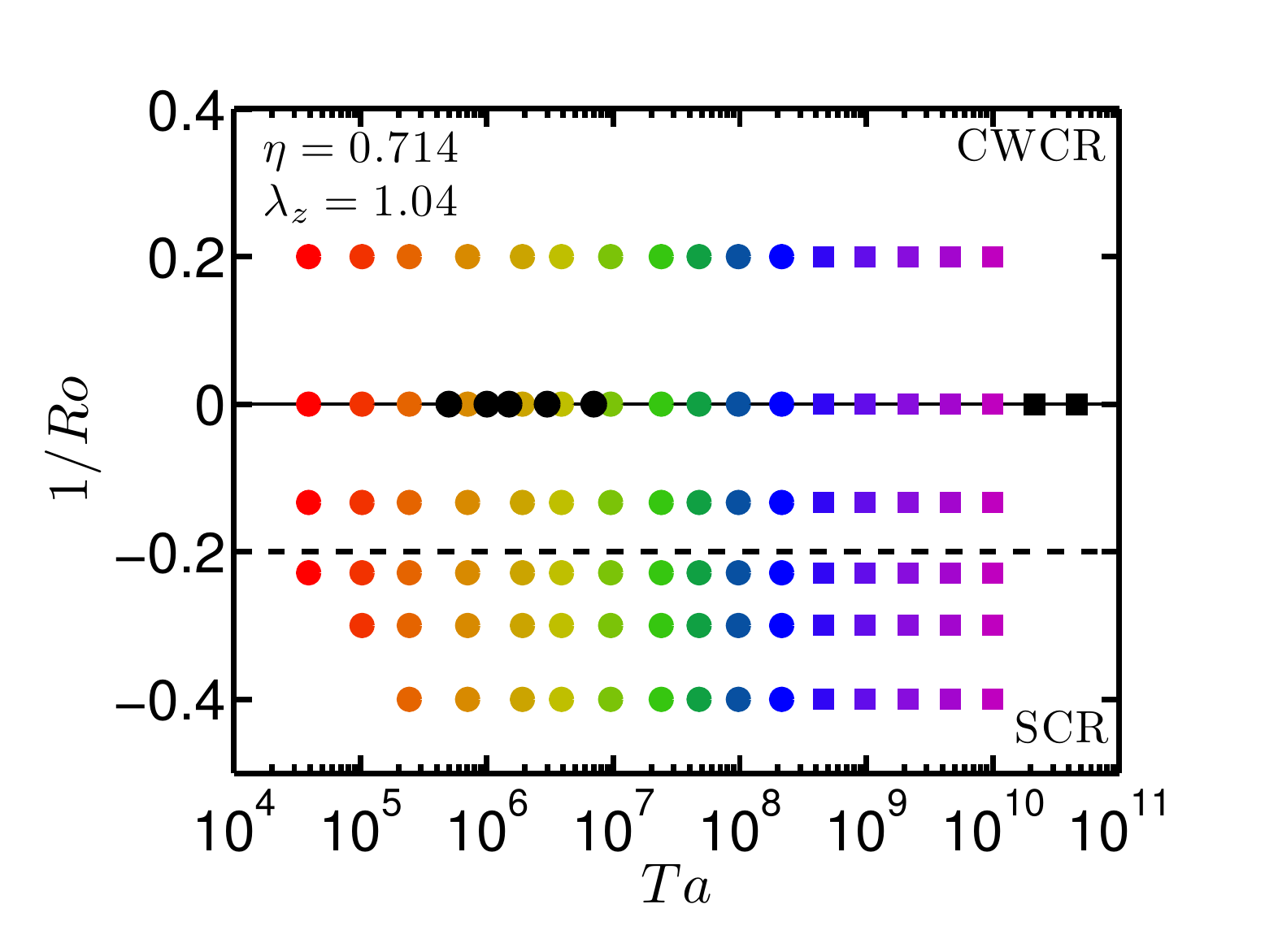}}\\
  \subfloat{\label{fig:TaEtaPhasespace}\includegraphics[width=0.49\textwidth,trim = 0mm 0mm 9mm 0mm, clip]{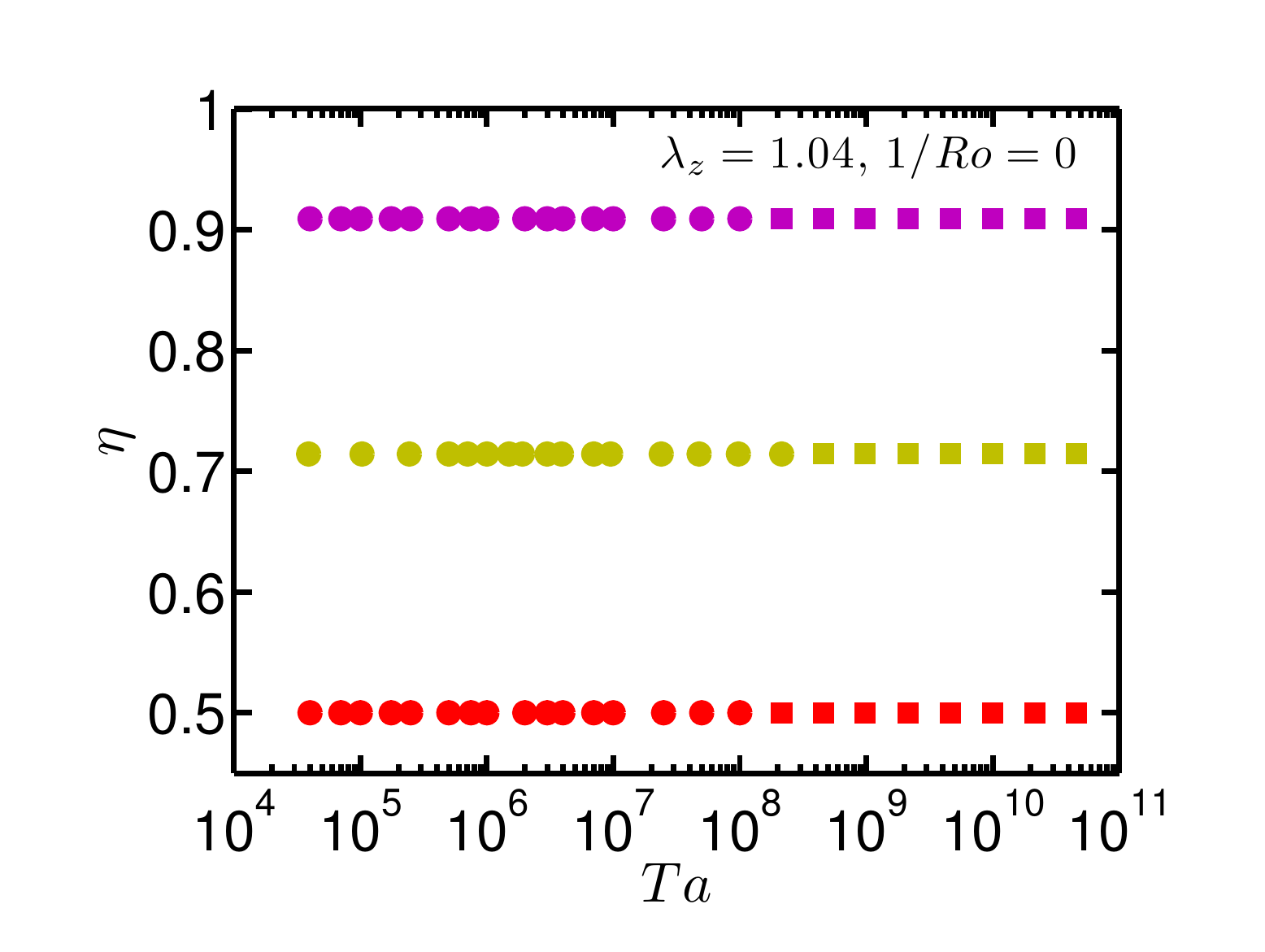}}
  \subfloat{\label{fig:TaLzPhasespace}\includegraphics[width=0.49\textwidth,trim = 0mm 0mm 9mm 0mm, clip]{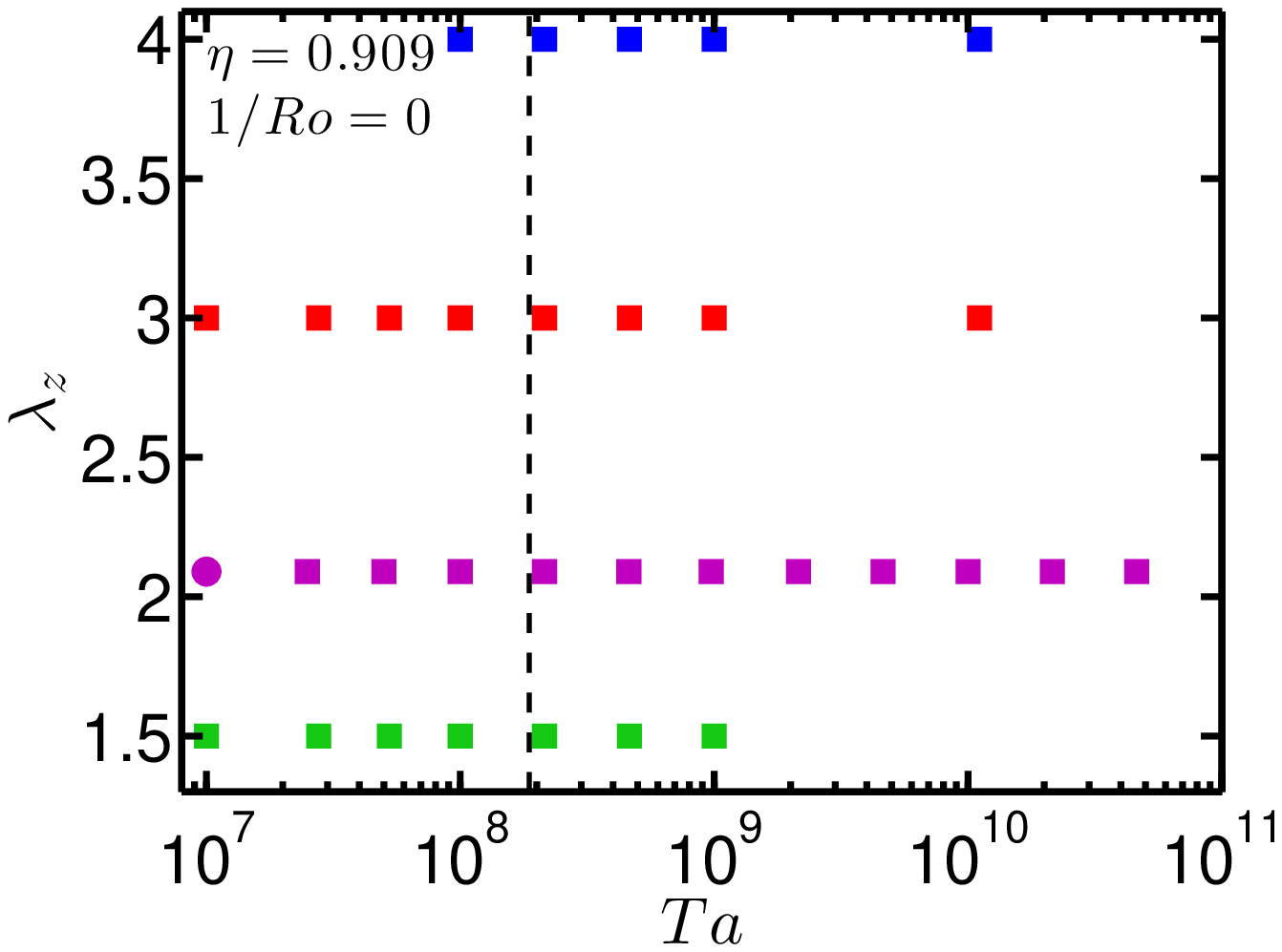}}\\
  \caption{Top left panel: Explored $(Re_i,Re_o)$ parameter space for $\eta=0.714$, $\lambda_z=1.04$. Top right panel: Same as left panel, but
now in the $(Ta,1/Ro)$ parameter space. In both panels the solid line indicates pure inner cylinder rotation, 
the dot-dash line indicates the Rayleigh stability criterium, 
while the dashed line indicates the asymptotic position of optimum transport in experiments, i.e. $\usro=-0.20$. The Rayleigh-stability line
lies outside the top right panel, at $\usro=0.83$. These lines divide the parameter space into the Rayleigh-stable zone, the co-rotating or weakly
counter-rotating regime (CWCR) and the strongly counter-rotating regime (SCR). 
Bottom left panel: explored $(Ta,\eta)$ parameter space for $\usro=0$, $\lambda_z=1.04$. Bottom right panel: explored $(Ta,\lambda_z)$ parameter space
for $\usro=0$, $\eta=0.909$. The dashed line indicates the cross-point of branches with different $\lambda_z$ in \cite{mar14}.
For the bottom panels, the same colour coding is maintained throughout the paper.
On all panels, circles indicate simulations of the ``full'' geometry, with three vortex pairs,
while squares indicate simulations of ``reduced'' geometries with forced rotational symmetry and one vortex pair.}
  \label{fig:Phasespace}
 \end{center} 
\end{figure}

A second--order finite--difference code was used with fractional time integration. The code was parallelized
using hybrid OpenMP and MPI-slab decomposition. Simulations were run on local clusters and on the supercomputer CURIE (Thin nodes) using a maximum of 8192 cores.
Details about the code can be found in \cite{ver96} and in \cite{ost13}. 
The explored parameter space from previous work \citep{ost13,ost14} was
extended through further simulations. Figure \ref{fig:Phasespace} shows the parameter space explored in this 
manuscript. Circles show simulations of a ``full'' geometry, i.e. a complete cylinder and with $\Gamma=2\pi$. Following the work of 
\cite{bra13}, the simulations with the largest $Ta$ were performed on ``reduced'' geometries to reduce computational costs, and these
are indicated as squares in the plots. The idea is that instead of simulating the whole cylinder, a cylinder wedge with rotational 
symmetry of order $n_{sym}$ is considered. The aspect ratio was also reduced to $\Gamma=2\pi/3$, 
accommodating a single vortex pair with the wavelength $\lambda_z=2\pi/3=2.09$. The vortical wavelength remains the same,
although there is a single vortex instead of the three vortex pairs having also the wavelength $\lambda_z=2\pi/3$. 
Other vortical wavelengths were also simulated using reduced geometries for $\eta=0.909$. 
We note that the aspect ratio $\Gamma$ is a geometrical control parameter, but $\lambda_z$ is a response of the system, which
depends both on $\Gamma$ and on the amount of vortex pairs which fit in the system. They are related by
$\lambda_z=\Gamma/n$, where $n$ is the amount of vortex pairs which fit in the system.
For all simulations axially periodic boundary conditions were used. Its consequences on the 
vortex wavelength are analyzed in section 4. Further details on the numerical resolution 
can be found in Table \ref{tbl:final} in the appendix.

\subsection{Explored parameter space}

The top two panels of figure \ref{fig:Phasespace} show the parameter space explored for $\eta=0.714$ in both $(Re_i,Re_o)$ and
$(Ta,1/Ro)$ to study the effects  of outer cylinder rotation. For $\eta=0.714$, reduced geometries simulate 
one sixth of the cylinder, i.e. $n_{sym}=6$ as used in \cite{ost14}. 
The chosen values of $\usro$ include a co-rotating outer cylinder ($\usro=0.20$), a weakly counter-rotating
outer cylinder ($\usro=-0.13$), counter-rotation near the asymptotic position of optimal transport, $\usro_{opt}$ 
($\usro=-0.22$), and two values of $\usro$ in the strongly counter-rotating regime ($\usro=-0.30$ and $\usro=-0.40$). 
No simulations were run in the Rayleigh-stable regime (i.e. when $r^2_o\omega_o>r^2_i\omega_i$)
as no evidence of turbulence was found in that regime up to $Ta\sim 10^{10}$ in \cite{ost14b}.

In addition, to study the effects of geometry, i.e. both the radius ratio $\eta$, and the vortical wavelength 
$\lambda_z$ (controlled through the aspect ratio $\Gamma$), additional simulations were performed. The bottom left panel 
shows that two additional radius ratios were simulated up to $Ta=4\cdot10^{10}$, one
with a larger gap ($\eta=0.5$) and one with a smaller gap ($\eta=0.909$). For $\eta=0.5$, one third of the cylinder 
($n_{sym}=3$) was simulated
for $Ta$ larger than $10^8$. This value of $n_{sym}$ for $\eta=0.5$ was shown not to affect the values of the
torque obtained in the simulations in \cite{bra13b}. 
For $\eta=0.909$, one twentieth of the geometry ($n_{sym}=20$) was used.

The bottom right panel shows the simulations with
varying vortical wavelength $\lambda_z$ done for $\eta=0.909$ and pure inner 
cylinder rotation. 
$\eta=0.909$ was chosen as we expect the effects of the coherent structures, and thus of $\lambda_z$, to be stronger 
for larger $\eta$ (see later sections \ref{sec:eta} and \ref{sec:lz} for an explanation). The values of $Ta$ 
simulated are around the range where \cite{mar14} have experimentally observed 
the crossing of different branches in $\nom(Ta)$ and also coincides with the onset of the ``ultimate'' regime.

\subsection{Non--dimensionalization}

The following non-dimensionalizations will be used: as the flow is simulated in a rotating frame, the outer 
cylinder is stationary,
and the system has an unique velocity scale, equal to $U\equiv r_i(\omega_i^{\ell}-\omega_o^{\ell})$ in the laboratory frame. 
All velocities are non-dimensionalized using $U$, i.e. $\tilde{\bu} = \bu/U$. The gap width $d$ is the characteristic length
scale, and thus used for normalizing distances. 

We define the normalized (non-dimensional) distance from the inner cylinder $\tilde{r} = (r-r_i)/d$ and the 
normalized height $\tilde{z}=z/d$. 
We furthermore define the time- and azimuthally-averaged velocity fields as:

\begin{equation}
 \bar{\tilde{\bu}}(r,z)=\langle \tilde{\bu}(\theta,r,z,t) \rangle_{\theta,t}~,
\end{equation}

\noindent where $\langle \phi(x_1,x_2,...,x_n) \rangle_{x_i}$ indicates averaging of the field $\phi$ with respect to $x_i$.  
As mentioned previously, the torque is non-dimensionalized as an angular velocity ``Nusselt'' number (EGL07), defined 
as $\nom=T/T_{pa}$, where $T_{pa}$ is the torque in the purely azimuthal flow. The torque is calculated from the radial derivative of $\langle\bar{\tilde{\omega}}\rangle_z$ at the inner and outer cylinders. 
The simulations are run in time until the respective values are equal within $1\%$. The torque is then taken as the average value of the inner and outer cylinder torques.
Therefore, the error due to finite time statistics is smaller than $1\%$.

From here on, for convenience we will drop the overhead tilde on all non-dimensionalized variables. 

\subsection{Structure of paper}

The organization of the paper is as follows. In section \ref{sec:outer}, we analyze the effect of rotating the outer cylinder. This is followed
by section \ref{sec:eta}, where we study the influence of $\eta$, and notice an analogy between the effects 
of smaller $\eta$ and larger $\usro$. 
In section \ref{sec:lz}, we consider the effects of the last parameter, the vortical wavelength $\lambda_z$.
We finish in section 6 with a summary of the results and an outlook for future work.

\section{The effect of outer cylinder rotation or the inverse Rossby number  dependence}
\label{sec:outer}

In this section we will study the effect of the Coriolis force ($\usro$), originating from 
the rotation of the outer cylinder, on the scaling of $\nom(Ta)$ with $Ta$ and, more
specifically, the effect of $\usro$ on the transition to the ultimate regime. 
Depending on the value of $\usro$, two distinct regimes will be identified: First a co- and weakly counter-rotating 
$Ro^{-1}$ range, denoted from here on as CWCR regime, and second the strongly counter-rotating $Ro^{-1}$ range, denoted 
from here on as SCR regime. 
The CWCR regime is found when the outer cylinder either is at rest, co-rotates with the inner cylinder, or 
slowly counter-rotates.
The counter-rotation must be slow enough such that no Rayleigh-stable zones are generated in the bulk of the flow. 
In this CWCR regime the Coriolis force is balanced through the bulk gradient of $\omega$. 
This can be derived from a large scale balance in the $\theta$-component of the velocity in equation (\ref{eq:rotatingTC}).
In summary, the non--linear term $u_r(\partial_r u_\theta + u_ru_\theta/r)$ 
and the Coriolis force term $-u_r\usro$ balance each other out \emph{on average} (cf. \cite{ost13} for the full derivation).
This results in a linear relationship between $\usro$ and $\partial_r \langle\bar{\omega}\rangle_z$ \citep{ost14c}. 

\begin{figure}
 \begin{center}
  \subfloat{\label{fig:TaNuas}\includegraphics[height=0.35\textwidth]{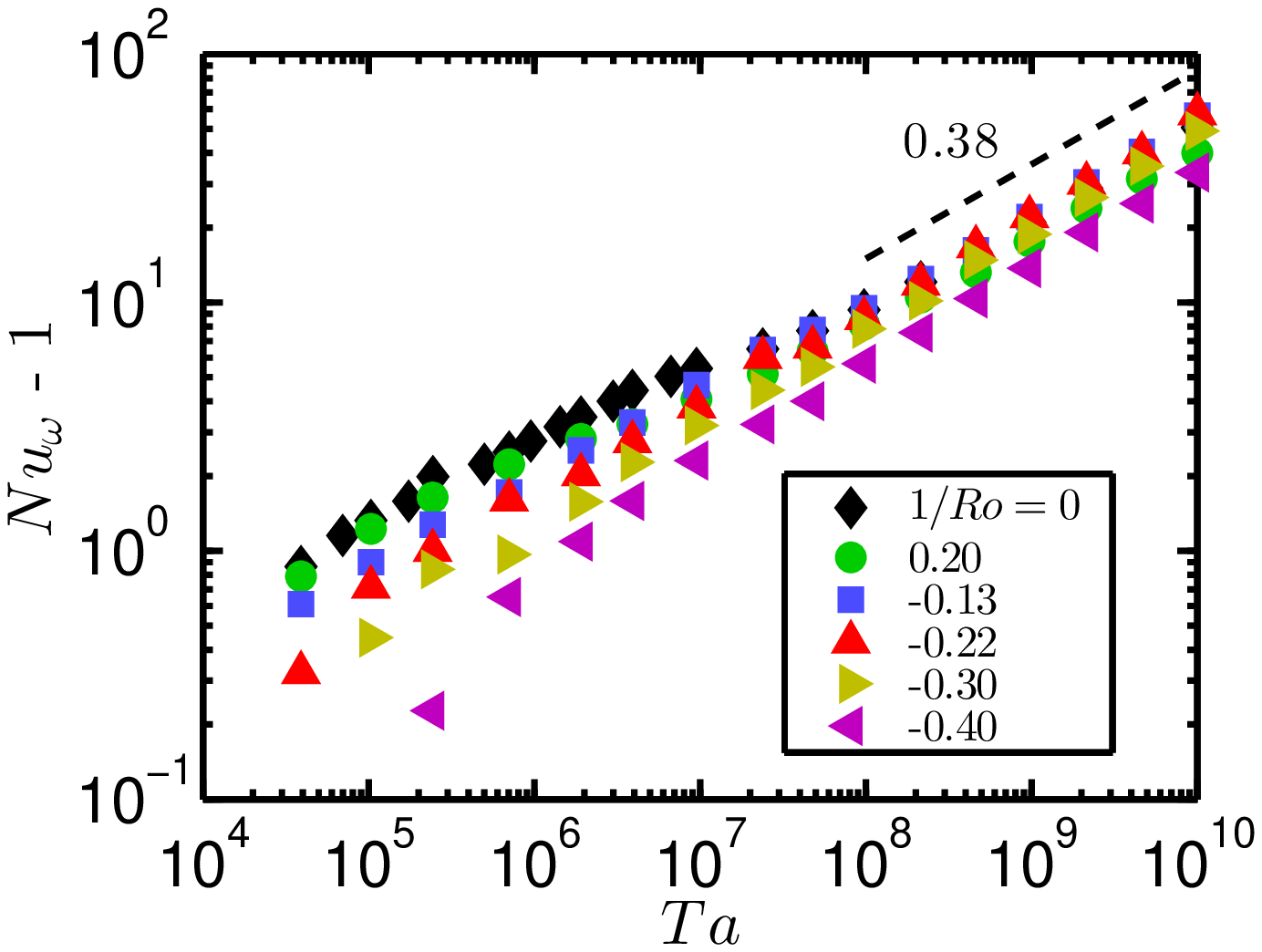}}                
  \subfloat{\label{fig:TaNuCompas}\includegraphics[height=0.35\textwidth]{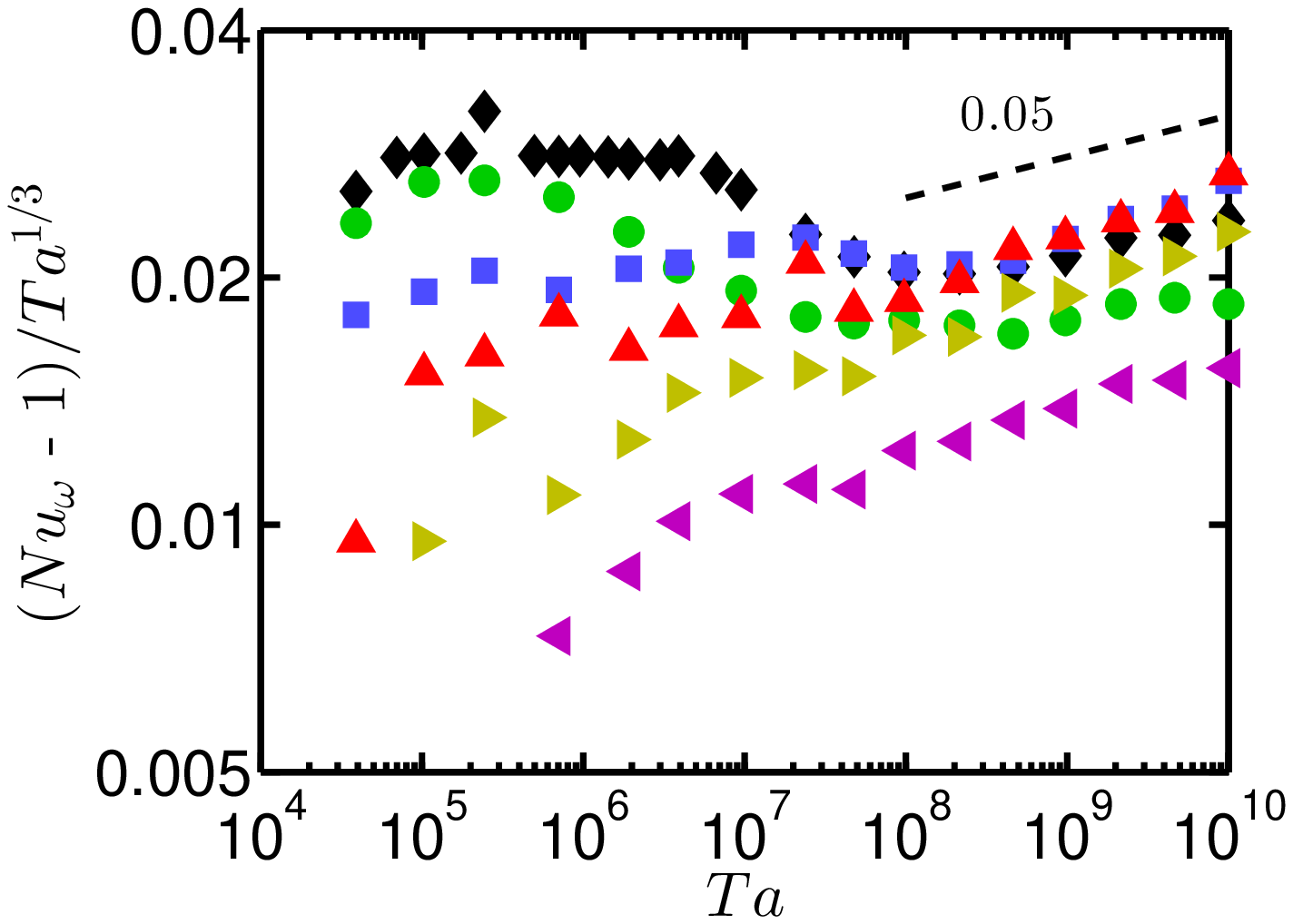}}\\
  \caption{Left panel: the non-dimensional torque $\nom-1$ versus the driving, i. e., the Taylor number $Ta$, for 
$\eta=0.714$ and six values of $\usro$. 
Right panel: the compensated Nusselt $(\nom-1)/Ta^{1/3}$ versus the driving $Ta$ for the same six values of $\usro$. 
The effective scaling law of $\nom\sim Ta^{0.38}$ is reached for all $\usro$ at the highest drivings $Ta$ beyond 
about $10^9$. However, the behavior in the classical regime $Ta$ less than $10^9$ depends heavily on $\usro$. Before 
the onset of the ultimate regime, we observe a transitional $Ta$-regime ranging from about $10^6$ to about $10^8$ associated
to the breakup of coherent structures for co-rotating and weakly counter-rotating cylinders ($-0.13\leq\usro\leq0.2$). 
For more positive values of $\usro$ this regime can be seen earlier, and is persistent for a larger $Ta$-range.
For the strongly counter-rotating cases ($\usro\leq-0.22$), an effective
 local scaling exponent with $\alpha > 1/3$ is seen in the classical regime. This can be
related to the interplay between Rayleigh-stable and unstable regions. }
  \label{fig:TaNuasfull}
 \end{center} 
\end{figure}

Taylor-Couette flow can be considered as being in the SCR regime, if the outer cylinder strongly counter-rotates 
and generates a Coriolis force which exceeds what the $\omega$-gradient can balance. The threshold value 
of $\usro$ corresponds to the flattest $\omega$ profile. This also is the value of $\usro$, 
for which $\nom(\usro)$ is found to be largest \citep{gil12,ost13}, denoted henceforth as $Ro^{-1}_{opt}$.
In this regime the turbulent plumes originating from the inner cylinder are not strong enough to overcome
the stabilizing effect of the outer cylinder, and the flow is divided into two regions, a Rayleigh-stable region
in the outer gap region, which plumes do not reach, and a Rayleigh-unstable region in the inner parts of the gap. For given 
Coriolis force, the relative sizes of these spatial regions depend on $Ta$, as for a stronger driving (i.e. larger $Ta$), 
the turbulence originating from the inner cylinder ``pushes'' these zones more towards the outer cylinder.
This may lead to switching between vortical states and jumps in global quantities as seen in \cite{ost13}.
The boundary between both regimes is at $\usro_{opt}$. Of course, $\usro_{opt}$ depends on $Ta$ too, 
due to effect of viscosity in the Coriolis force balance \citep{ost13}, and only saturates
to $\usro_{opt}(Ta\to\infty)=-0.20$ for sufficiently high drivings of $Ta\sim 5\cdot 10^8$ and more (cf. both panels of figure \ref{fig:TaNuasfull} and \cite{bra13}).

Figure \ref{fig:TaNuasfull} shows both $\nom-1$ and the compensated Nusselt 
number $(\nom-1)/Ta^{1/3}$ versus $Ta$ for $\eta=0.714$ and the six values of $\usro$ studied. 
For the largest drivings (i.e. $Ta>10^9$) 
all values of $\usro$ reach the effective scaling law $\nom\sim Ta^{0.38}$ (with a different
amplitude), similar to what was reported in the experiments by \cite{gil11}.
However, very different behavior can be seen for $Ta<10^9$, i.e. before the onset of the ultimate regime.

In the CWCR regime ($\usro \geq \usro_{opt} = -0.20$),
the Coriolis force is reflected in the flow structure through
the bulk gradient of $\omega$, making it either flatter as in the case of weak-counter rotation, or steeper, 
as in the case of co-rotation (if the driving is sufficiently large). 
A consequence of the angular velocity gradient in the bulk is that
large scale structures can be weakened or even completely dissapear in the CWCR regime.
These changes in $\omega$-gradient strongly affect the capability of plumes to ``coordinate'' and form a large-scale wind,
which in turn leads to an earlier (or later) onset of the sharp decrease in the local exponent $\alpha$
in the scaling law $(\nom-1)\sim Ta^{\alpha}$  associated 
with the breakdown of coherence, and the onset of time dependence in $\nom$ \citep{ost14}.

For the case of co-rotating cylinders ($\usro=0.20$), this happens when the system enters 
the so-called ``wavelet'' regime, characterized by moving waves in the boundary regions 
between a pair of Taylor vortices  \citep{and83,and86}. These waves move with different speeds, and as a consequence
this regime is not stationary in any reference frame. This regime only persists for a small range of $Ta$, and eventually
all remnants of Taylor vortices vanish. Axial dependence of the flow structure is almost completely lost, 
even at $Ta$ as low as
$Ta\approx 5\cdot10^7$. Unlike the case of $\usro=0$ studied in \cite{ost14}, however, in this transitional regime, 
the large--scale rolls already completely vanished, but for $\usro=0.20$ this does not immediately lead to the transition to the ultimate regime. 
After its sharp decrease, $\alpha$ does not exceed $1/3$. Instead, at a driving strength around $Ta\approx 10^7$ (coinciding with
the disappearance of the structures),  the local effective scaling exponent $\alpha$ has increased 
to $\alpha\approx1/3$, and then stops growing.
Only if $Ta$ increases further and the shear in the boundary layers grows past a threshold, a shear-instability takes place,
and the system transitions to the ultimate regime.

For the case of counter-rotating cylinders, (i.e. $\usro<0$), $\alpha$ 
can locally grow beyond $\alpha = 1/3$ in the classical regime. This is unexpected, as values of $\alpha$ 
larger than one third 
have been associated to the transition to turbulence of the boundary layers in the
context of both Rayleigh-B\'enard convection \citep{he12}, and TC with a stationary outer cylinder \citep{ost14}.
However, in this case, the shear in the boundary layers is too low so the boundary layers still stay laminar. 

For counter-rotating cylindres, a wide range of flow configurations is available
in the low-$Ta$ regime \citep{and86}. We can relate local steps in $\alpha$ to the switching between such 
flow configurations. The interplay between Rayleigh-stable and -unstable regions can also play a role. Larger 
drivings cause the Rayleigh-unstable region to grow, and thus to increase the transport.
These two effects lead to larger increases in the non-dimensional torque than what is expected for pure inner cylinder
rotation, and explain the large values of $\alpha$ seen. 

\begin{figure}
 \begin{center}
  \subfloat{\label{fig:q1meEta0714am02}\includegraphics[height=0.40\textwidth,trim = 40mm 0mm 40mm 0mm, clip]{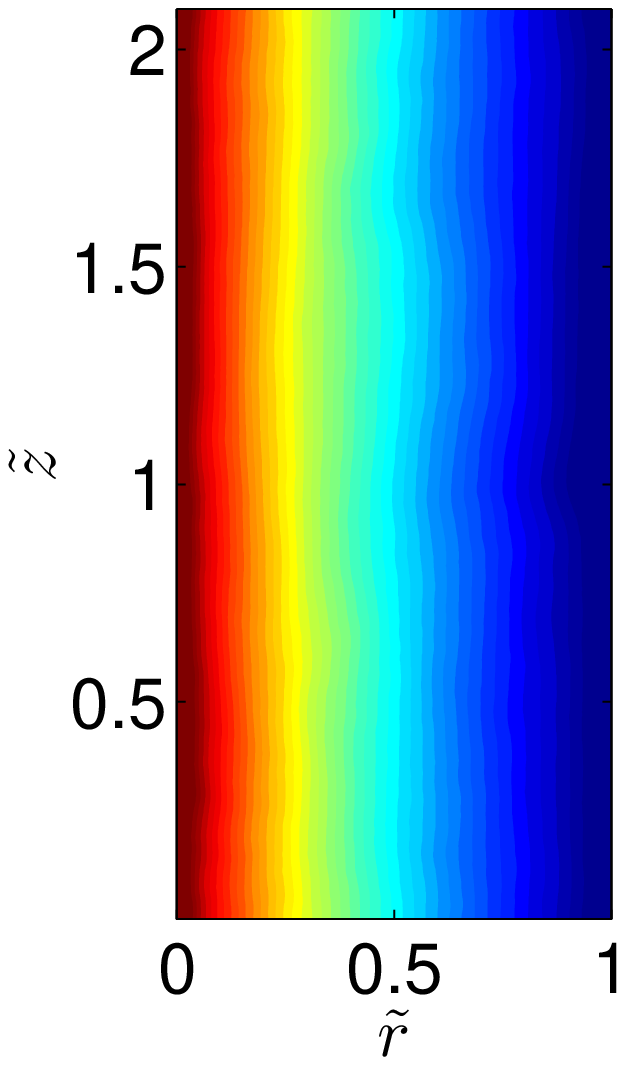}}
  \subfloat{\label{fig:q1meEta0714a0}\includegraphics[height=0.40\textwidth,trim = 40mm 0mm 40mm 0mm, clip]{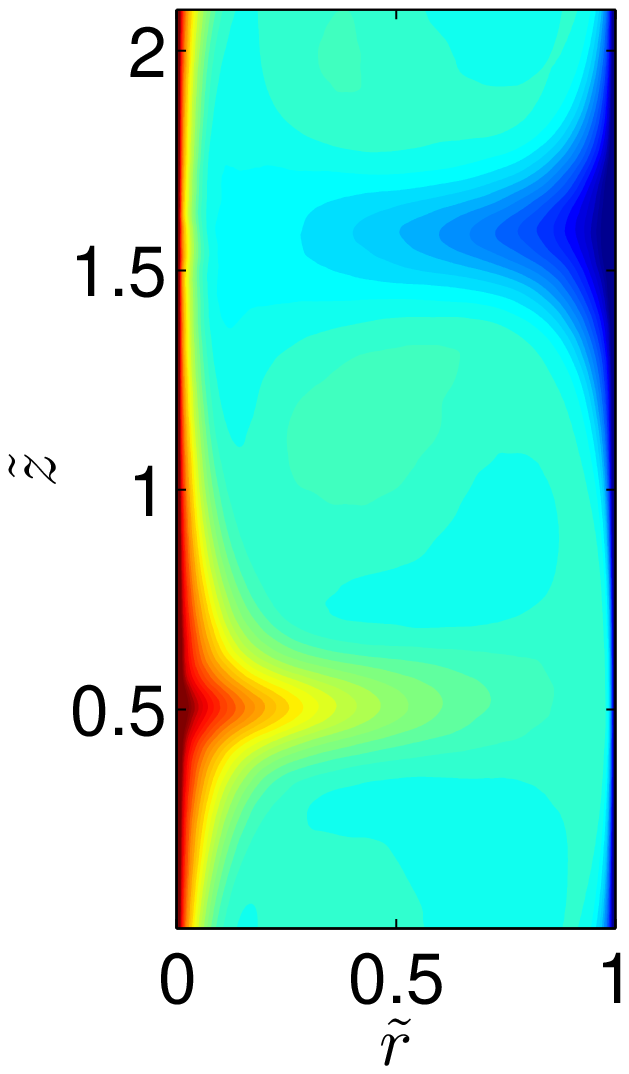}}
  \subfloat{\label{fig:q1meEta0714a04}\includegraphics[height=0.40\textwidth,trim = 30mm 0mm 40mm 0mm, clip]{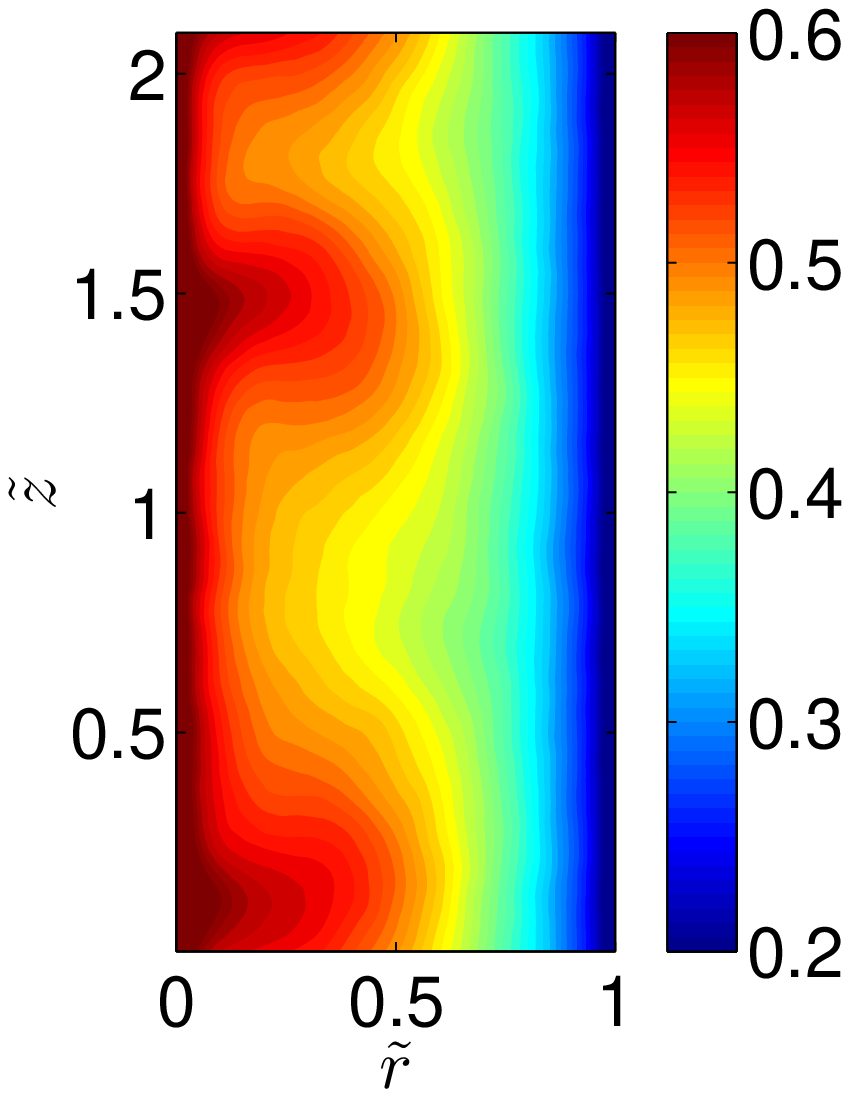}}
  \caption{Contour plots of the azimuthally- and time averaged angular velocity field $\bar{\omega}$ for $Ta = 10^{10}$,
$\eta=0.714$ and three values of $\usro$. The left panel corresponds to $\usro=0.2$ (CWCR regime) and shows no traces 
of axial dependence. Plumes detach rapidly into the bulk, mix there strongly, and thus cannot form large-scale 
structures. The middle panel corresponds to 
$\usro=-0.22$ ($\approx \usro_{opt}$). The reduced plume mixing enables the formation of large-scale structures, 
and a strong signature of them can be seen in the averaged angular velocity field. The right panel
corresponds to $\usro=-0.40$ (SCR regime) and also shows some signatures of large-scale structures. 
However, these do not fully penetrate the gap but stop at the border to the Rayleigh-stable zones near the outer cylinder.}
  \label{fig:OmegaFullEta0714as}
 \end{center} 
\end{figure}

\begin{figure}
 \begin{center}
  \includegraphics[width=0.47\textwidth]{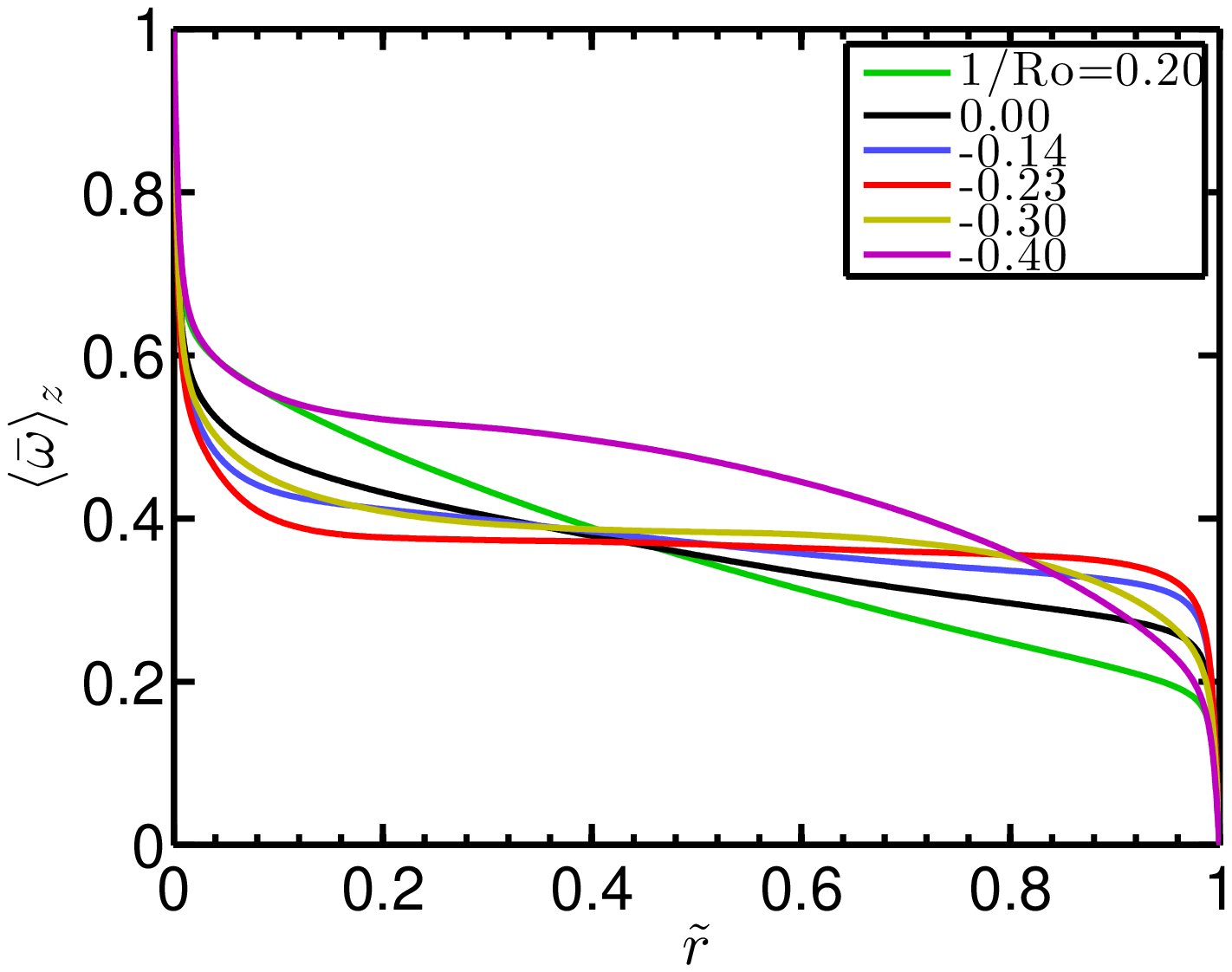}
  \includegraphics[width=0.47\textwidth]{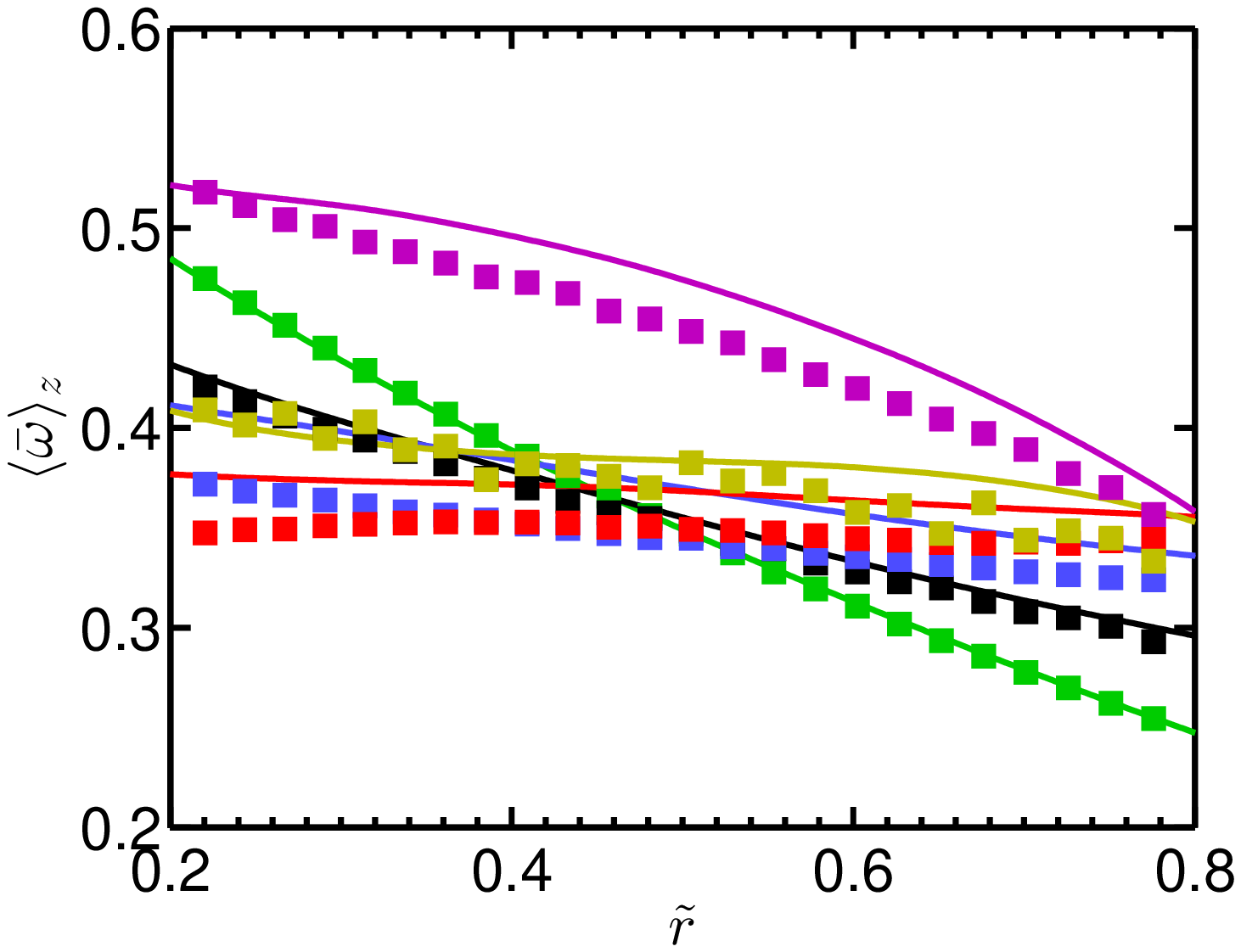}
  \caption{Azimuthally, axially and time-averaged and non-dimensionalized angular velocity 
profiles $\langle\bar{\omega}\rangle_z$ for $\eta=0.714$, $Ta=10^{10}$, and six values of $\usro$. 
For co- and weakly counter-rotating cylinders, we see that the bulk $\bar{\omega}$ profiles become flatter as $\usro$ becomes 
more negative. Thus, the angular velocity difference, which the plumes encounter when detaching from the BL
and entering the bulk, is larger for 
more negative $\usro$. The right panel shows a zoom-in
in the bulk region of the left panel. LDA data from experiments from \cite{gil12}, for which $Re_i-Re_o=10^6$
have been superimposed. Note the good 
agreement between both datasets for values of $\usro$ in the CWCR regime, while discrepancies
exist for values of $\usro$ around the optimum and in the SCR regime. This is attributed to 
the axial dependence of the profiles, which exists in this regime, see figure \ref{fig:OmegaFullEta0714as},
and as experimental data is measured at \emph{fixed} height, while numerical data are axially 
averaged. }
 \label{fig:OmegazEta0714as}
 \end{center} 
\end{figure}

To further illustrate the effect of the Coriolis force on the large-scale structures, 
figure \ref{fig:OmegaFullEta0714as} presents a contour plot of $\bar{\omega}$ 
in the CWCR regime $\usro=0.20$, around the optimum $\usro=-0.22\approx\usro_{opt}$ and in the SCR regime $\usro=-0.40$.
Figure \ref{fig:OmegazEta0714as} shows the axially-averaged angular velocity profiles $\langle \bar{\omega} \rangle_z$
for $\eta=0.714$ and the six values of $\usro$ simulated here.
The large--scale structures cannot be seen in the left panel of figure 
\ref{fig:OmegaFullEta0714as}, which corresponds to $\usro=0.20$ 
(co-rotating cylinders), but they are pronounced for the other two panels ($\usro=-0.22$ and $\usro=-0.40$). 
As shown in figure \ref{fig:OmegazEta0714as}, in the CWCR regime, the bulk sustains a large $\bar{\omega}_z$ gradient, 
and to accomodate for this, there is smaller $\bar{\omega}_z$ jump across the boundary layers. 
Plumes ejected from both cylinders can now mix easier when entering the bulk. 
As a consequence, the large-scale structures, which essentially consist of unmixed plumes, 
break up easier and thus do that for lower values of $Ta$. For this reason they have completely vanished
in the left panel of figure \ref{fig:OmegaFullEta0714as}. 

If we now decrease $\usro$, the profile becomes flatter. The effect of this
is visible in the middle panel of Figure \ref{fig:OmegaFullEta0714as} 
showing $\bar{\omega}$ for $\usro=-0.22$. It can be seen 
from figure \ref{fig:OmegazEta0714as} that this value of $\usro$ corresponds to the flattest $\omega$-profile available, 
and it is also the closest to the experimental optimum transport $\usro_{opt}(Ta\to\infty) = -0.20$. 
A very marked signature of the large-scale structure on $\bar{\omega}$ can be seen. 
This is because a very flat $\bar{\omega}$ profile will sustain a large $\bar{\omega}$ jump across the boundary layer,
and thus plumes detach less violently into the bulk, thus stabilizing the large-scale structures.
Therefore, we can relate the flatness of the $\bar{\omega}$-profile
to the strength of the large-scale circulation, and this in turn can be related to the optimum in $\nom(\usro)$.
As mentioned in \cite{bra13b}, optimum transport coincides with the strongest mean circulation.
Plumes travel faster from one cylinder to the other when the large-scale circulation is strongest, and thus
more angular momentum is transferred. 
We also highlight that the signature of the large-scale structures on the mean
azimuthal flow remains even in the ultimate regime, and is also seen in
experiment at $Ta\sim10^{12}$ \citep{hui14}. 
Thus in general the vanishing of the rolls appears to be independent from the transition to the ultimate regime.
Only in the special case of pure inner cylinder rotation these two effects coincidentally occur at the same $Ta$.

In the right panel of figure \ref{fig:OmegaFullEta0714as}, we can see that once the Coriolis force is
sufficiently large, the vortices cannot fully penetrate the domain. Near the outer cylinder, the flow is 
predominantly Rayleigh-stable.
Rayleigh-stable zones are well mixed, as transport here happens through intermittent turbulent bursts, 
instead of convective transport by plumes and vortices \citep{bra13b}.
Thus, in Rayleigh-stable regions, no rolls can be seen in the averaged fields.
The effect of the neutral surface can also be observed in the averaged $\omega$ profiles (cf. figure \ref{fig:OmegazEta0714as}).
The two simulated cases in the SCR regime, ($\usro=-0.30$ and $\usro=-0.40$) show an outer cylinder boundary layer 
which with more and more negative $\usro$ extends deeper into the flow, and the distinction from the bulk is blurred away.

\begin{figure}
 \begin{center}
  \includegraphics[width=0.47\textwidth]{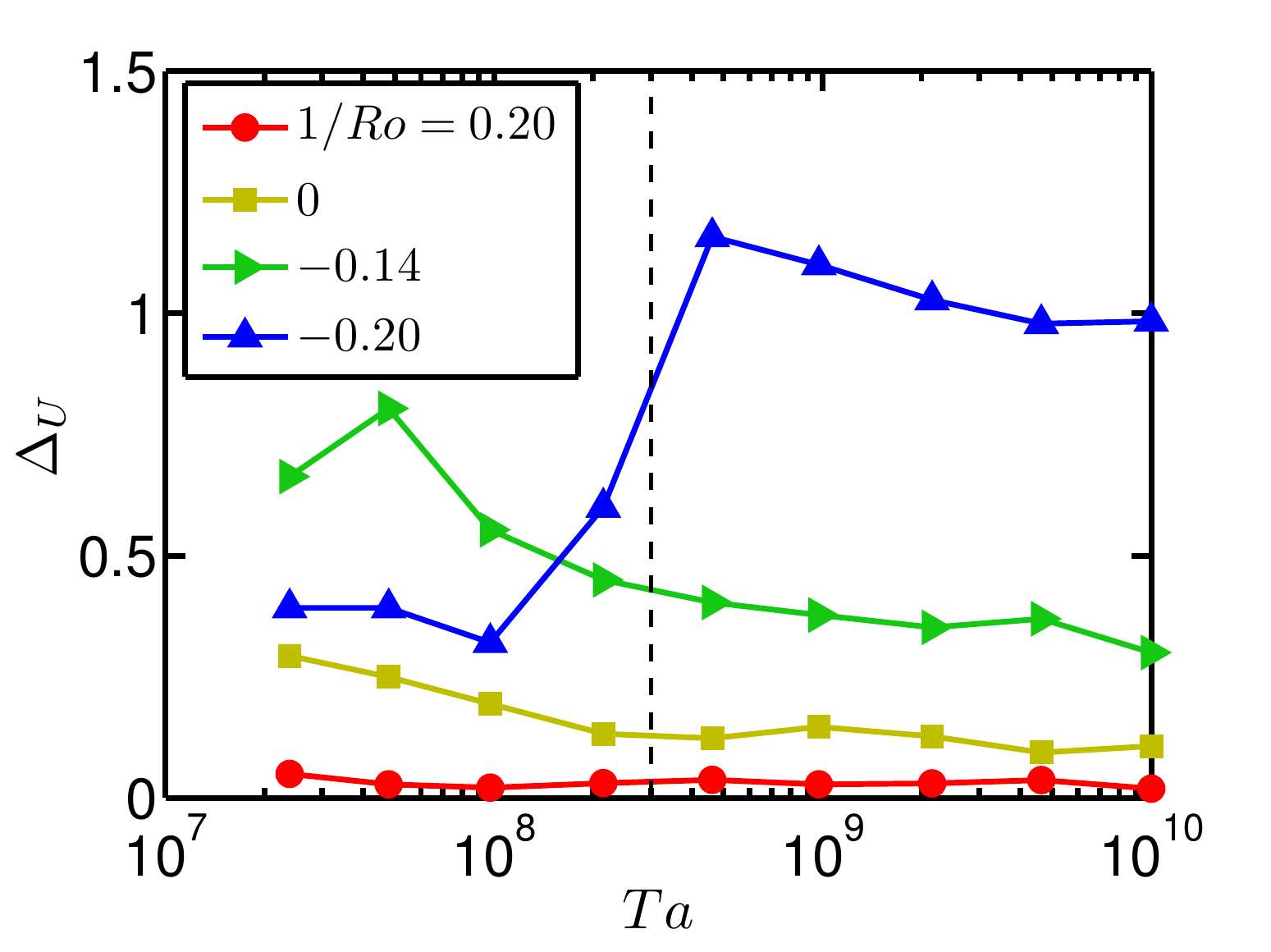}
  \caption{The axial velocity spread measure $\Delta_U$ versus $Ta$ for the four values of $\usro$ in the CWCR regime. 
The dashed line indicates the approximate value of $Ta$ where the flow transitions to the 
ultimate regime for all values of $\usro$, which was 
previously associated with the vanishing of the large-scale
structures. For co-rotating cylinders ($\usro=0.20$), at $Ta$ as low 
as $Ta\approx 10^7$ no axial dependence is seen, well before the transition.
For counter-rotating cylinders a sharp increase of the axial velocity spreading measure $\Delta_U$ can be seen, 
which then slowly decreases with increasing $Ta$. The sharp increase in $\Delta_U$ can be associated to the 
growth of the Rayleigh-unstable zones. For low $Ta$, the mid-gap is in a Rayleigh-stable zone mixed by bursts,
while for large $Ta$, the mid-gap is in a Rayleigh-unstable zone, dominated by rolls leading to a strong height dependence. The large axial spreads explain the discrepancies when comparing (axially averaged) DNS data to experimental data measured at a fixed height.}
 \label{fig:deltauomegaplus}
 \end{center} 
\end{figure}

To further disentangle the effect of axial dependence and the transition to the ultimate regime we show the loss of axial 
dependence characterized by a special spread measure $\Delta_U$ as a function of the driving $Ta$ in figure \ref{fig:deltauomegaplus}.
$\Delta_U$ is defined as $\Delta_U=(\max_z(\bar{u_\theta}(r_a,z))-\min_z(\bar{u_\theta}(r_a,z)))/\langle\bar{u_\theta}(r_a,z)\rangle_z$,
with $r_a$, the mid-gap, defined as $r_a=r_i+d/2$, the arithmetic mean of the inner and outer cylinder radii.
 When measuring the axial spread, the velocity is averaged in time, and azimuthally, as the flow is homogeneous in the azimuthal direction.
As stated previously, for co-rotating cylinders, the axial dependence disappears for low drivings 
corresponding to those in the transitional regime, and associated to the appearance of the ``wavelet'' states.
For counter-rotating cylinders, a sharp jump in $\Delta_U$ can be noticed. This is due to 
$\Delta_U$ being measured at the mid-cylinder $\tilde{r}=\tilde{r}_a$.
For low drivings, $\tilde{r}_a$ is located in the Rayleigh-stable zones, and the flow is mixed better. 
As the driving increases, 
turbulence from the inner cylinder pushes the neutral surface, which divides the stable and unstable zones
further towards the outer cylinder. As a consequence of this pushing, $\tilde{r}_a$ 
is no longer in the Rayleigh-stable zone, but instead in the Rayleigh-unstable zone. This
zone is dominated by large-scale structures. This makes the axial dependence increase and 
provides more evidence that the vanishing of the Taylor-rolls is only coincidental with the transition to the 
ultimate regime for pure inner cylinder rotation.

\begin{figure}
 \begin{center}
   \includegraphics[width=0.98\textwidth]{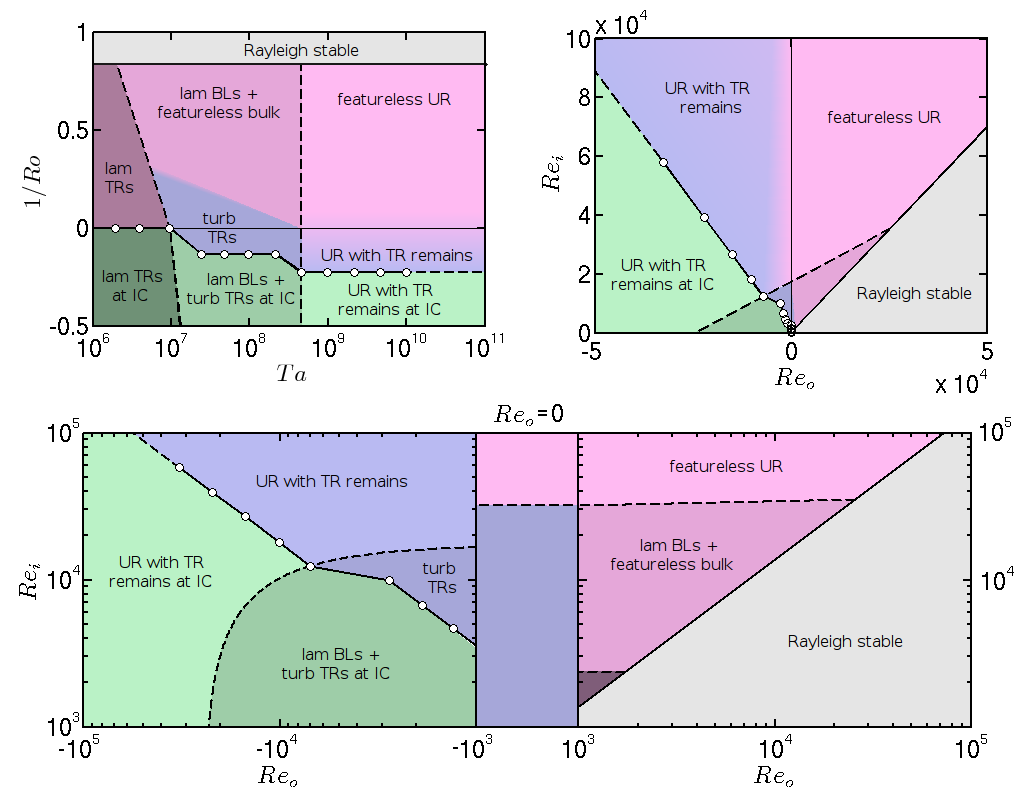}
    \caption{Transition between different regimes in the $(Ta,\usro)$ (top-left) 
and $(Re_i,Re_o)$ (top-right and bottom) parameter spaces for $\eta=0.714$.
The hollow circles indicate the location of optimal transport, and serve as an indication 
of the movement of the division between CWCR (blueish and reddish) and SCR (greenish) 
regimes with $Ta$. In both DNS and experiments, 
$\usro_{opt}$ reaches an asymptotic value for $Ta>5\cdot10^8$. For larger $\eta$ (smaller gap), this
separation line moves towards smaller $\usro$. For $Ta\lesssim 10^7$, we have the rich  
variety of different states of \cite{and86}, not detailed in this diagram.  This region appears explicitly in the top-left panel as ``lam TRs'' and ``lam TRs at IC'', but is not shown in the other two due to the axes used. Abbreviations: boundary layer (BL), Taylor rolls (TR), ultimate regime (UR), and inner cylinder (IC). }
   \label{fig:phasespacenew}
 \end{center} 
\end{figure}

As mentioned previously, the value of $\usro_{opt}$, and thus of the border between 
the CWCR and the SCR regimes depends on $Ta$. This is summarized in figure \ref{fig:phasespacenew}, which shows the approximate
division between the different flow regimes explored in this paper in both the $(Ta,\usro)$ and the $(Re_i,Re_o)$ parameter spaces, both
for $\eta=0.714$.  
$\usro_{opt}$, and thus the division between the regims can be seen to saturate for $Ta\sim 5\cdot10^8$, when driving is large enough, and
the mean $\bar\omega(r)$ profile at $\usro_{opt}$ is completely flat.

Finally, to further justify the division of the flow into the CWCR 
and the SCR regimes with decreasing inverse Rossby number $Ro^{-1}$, we can quantify the distribution of Rayleigh-stable 
and unstable zones as a function of $\usro$. This is done by looking at the PDF of $\tilde{r}_N$, i.e.
the collection of points outlining the neutral surface $\tilde{r}_N=\tilde{r}_N(t,\theta,z)$.  This is, the 
border between Rayleigh-stable outer gap range and the Rayleigh-unstable inner gap parts, and given as 
the points for which $\omega(t,\theta,z,\tilde{r}_N)=0$
in the laboratory (non-rotating) frame.  For counter-rotating cylinders, the neutral surface defines the instantaneous border between Rayleigh-stable and Rayleigh-unstable zones. For co-rotating cylinders, the neutral line does not exist, and the whole flow is either Rayleigh-stable or Rayleigh-unstable. In principle, the neutral surface might be fragmented, and thus the position of $\tilde{r}_n$ multivalued. However, this is usually not the case. When taking the ensemble, all values are considered, as this does not change the PDFs significantly.

\begin{figure}
 \begin{center}
   \includegraphics[width=0.47\textwidth]{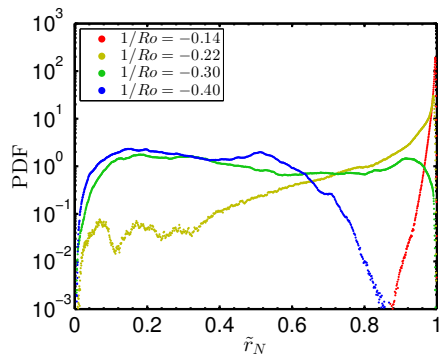}
    \caption{DNS results for the PDFs of the radial position $\tilde{r}_N$ of the neutral surface, at the border between Rayleigh-stable
and Rayleigh-unstable regions, for the four simulated negative values of $\usro$ 
(i.e. for counter-rotating cylinders) for $Ta=10^{10}$. For $\usro=-0.14$ (CWCR) and $\usro=-0.22$ (close to the optimum), the PDFs show that 
the destabilizing action of the inner cylinder causes the 
Rayleigh-stable regions to be confined only very closely to the outer cylinder. For $\usro=-0.22$ we can begin to see 
a limited amount of Rayleigh-stable zones in the whole domain, as $-0.22$ is slightly more negative than $\usro_{opt}$. For $\usro=-0.30$ (SCR), 
the stabilization due to the Coriolis force increases, and the border between the regions can be anywhere in 
the gap, indicating a mixture of stable and unstable zones everywhere in the gap. Finally, for $\usro=-0.40$ (also SCR), 
the border between both zones never gets close to the outer cylinder. For this case, the portion of the gap width with $\tilde{r} > 0.84$ 
is always Rayleigh-stable. 
}
   \label{fig:nlpdf}
 \end{center} 
\end{figure}

Figure \ref{fig:nlpdf} shows the PDFs of $\tilde{r}_N$ calculated for the four 
negative values of $\usro$ at the largest driving simulated here. 
The difference between the two regimes can clearly be noticed. 
In the CWCR regime and near the optimum, the border between the zones is located very closely to the outer cylinder, 
which means that almost all the domain is Rayleigh-unstable and dominated by plumes or rolls. In the SCR regime, 
the border between 
the zones is pushed closer towards the inner cylinder, and Rayleigh-stable
zones appear all over the gap. For the most negative simulated value of $\usro$, i.e. $\usro=-0.40$, the areas 
near the outer cylinder
are permanently Rayleigh-stable, and transport occurs in intermittent bursts which mix this zone well. 
This causes the partial dissappearance of axial dependence seen in the right panel of Figure \ref{fig:OmegaFullEta0714as}.

\section{The effect of radius ratio or the $\eta$-dependence}
\label{sec:eta}

In the previous section we showed that for $\eta=0.714$ the transition to the ultimate regime and the vanishing
of the rolls only (incidentally) co-occur at the same $Ta$ 
for pure inner cylinder rotation. Flatter bulk $\omega$-profiles result in stronger
large-scale structures, and steeper bulk $\omega$-profiles result in weaker large-scale structures which
vanish at $Ta\sim 10^6$. Now we will show that we can modify the $\bar\omega(r)$ profile in the bulk not only by varying 
the Coriolis force, but
also by changing the radius ratio $\eta$ (or the gap width). In this section, we will thus analyze the influence of $\eta$, 
to understand whether the co-ocurrence of the vanishing large scales and the boundary layer transition observed
for pure inner cylinder rotation is just a coincidence seen in the case $\eta=0.714$.

\begin{figure}
 \begin{center}
  \subfloat{\label{fig:TaNuEtas}\includegraphics[height=0.35\textwidth,trim = 0mm 0mm 0mm 0mm, clip]{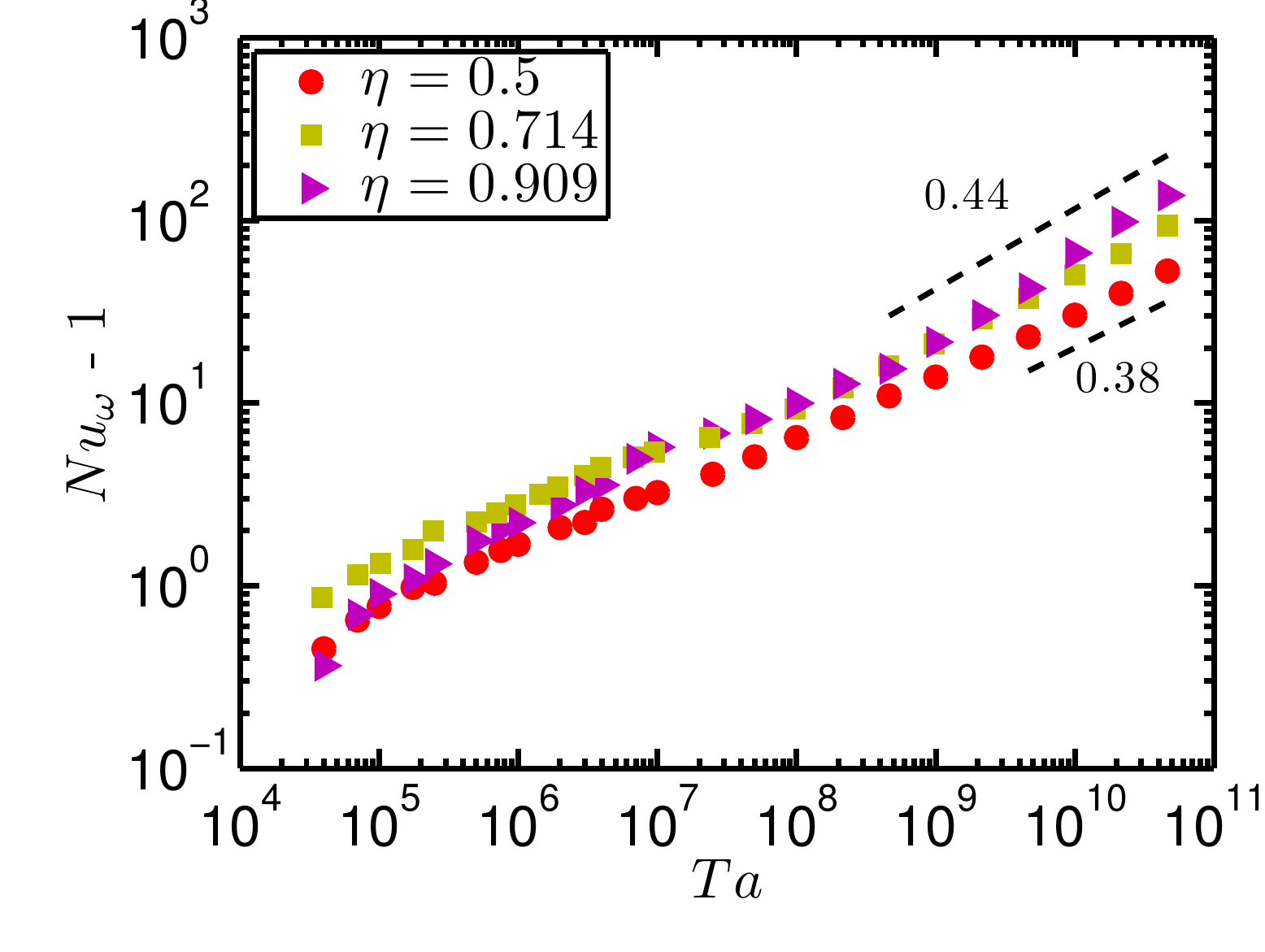}}                
  \subfloat{\label{fig:TaNuCompEtas}\includegraphics[height=0.35\textwidth,trim = 0mm 0mm 0mm 0mm, clip]{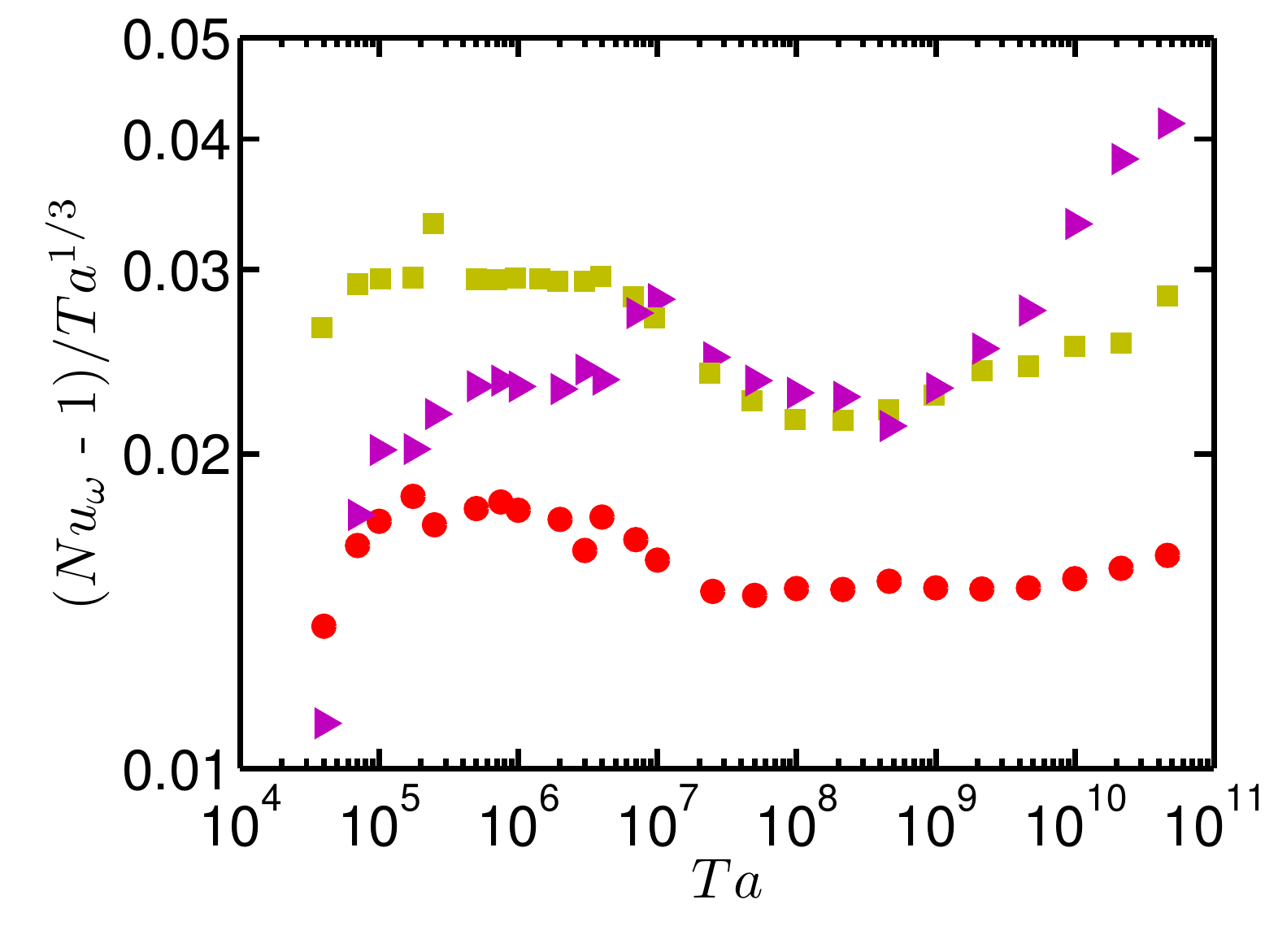}}\\
  \caption{Left panel: The nondimensional torque $\nom-1$ versus driving $Ta$ for pure inner cylinder rotation 
$\usro =0$ and three values of $\eta$.  Right panel: the compensated Nusselt $(\nom-1)/Ta^{1/3}$ versus driving strength  
$Ta$ for the same three
values of $\eta$. The asymptotic effective scaling laws of the ultimate regime are reached for all values of $\eta$ at large enough
drivings. For $\eta=0.909$ jumps in $\nom(Ta)$ can be seen for the highest drivings (around $Ta\sim10^{10}$). These jumps cause
the exponent of the local scaling laws to be around $0.44$, and are caused by changes in the large scale structures.
}
  \label{fig:TaNuEtasfull}
 \end{center} 
\end{figure}

Figure \ref{fig:TaNuEtasfull} shows both the Nusselt number and the compensated Nusselt number 
plotted as a function of $Ta$ for the three values of $\eta$ simulated.
As seen in \cite{ost14c} for $\eta=0.714$ (and now also for $\eta=0.909$), 
the flow undergoes a structural transition at around $Ta\approx 3\cdot10^6$, 
where the local exponent $\alpha$ of the 
effective scaling law $Nu\sim Ta^\alpha$ rapidly decreases. This is associated with the breakdown of coherence in the flow 
and the onset of time-dependence in the Nusselt number.
For $\eta=0.714$ and $\eta=0.909$, the effective exponent $\alpha$ begins to increase again after
this breakdown. We can 
say that the flow transitions 
to the ultimate regime once $\alpha>1/3$, and this happens at about $Ta\approx 3\cdot10^8$. 
This $Ta$ value coincides with the experimentally observed
value for the transition to the ultimate regime for $\eta=0.909$, cf. \cite{rav10}.

For $\eta=0.5$ a different behavior can be seen. After the breakdown of coherence, the transitional regime 
with $\alpha\approx1/3$ goes on for three decades in $Ta$, up to $Ta\approx 10^{10}$ (last three data points of the panel). 
An increase in $\alpha$ only happens for the last three data points, with $Ta>10^{10}$. 
This might be the beginning of the transition to the ultimate regime, observed at about that value of $Ta$
in the experiments by \cite{mer13}.
We emphasize that the behavior of the $\nom(Ta)$ curve for $\eta=0.5$ is very similar to the one 
seen for $\eta=0.714$ and $\usro=0.20$ (cf. figure \ref{fig:TaNuasfull}), while the $\nom(Ta)$ curve for $\eta=0.909$
is similar to the one for $\usro=-0.14$ and $\eta=0.714$.

We thus can draw an analogy between the effects of varying $\eta$ and those of changing $\usro$.
The larger the gap or the smaller $\eta$ is, the more the flow feels the curvature. This is reflected in an asymmetry 
between inner and outer cylinder, since the inner cylinder curvature becomes increasingly stronger relative 
to the outer cylinder curvature. Also the exact relationship 
$\eta^{-3} \partial_r \langle \omega \rangle |_o =  \partial_r \langle \omega \rangle |_i$ (cf. \cite{gil12})
must hold in both 
boundary layers due to the $r$-independence of the angular velocity current 
$J^{\omega}=r^3(\langle u_r\omega\rangle_{z,\theta,t}-\nu\partial_r\langle\omega\rangle_{z,\theta,t})$ (EGL07). 
For $\eta=0.5$ we have $\eta^{-3}=8$ 
and the $\omega$-slope at 
the inner cylinder is eight-fold steeper than the outer cylinder $\omega$-slope. Thus the 
inner-outer asymmetry is expected to become much more dominant for $\eta = 0.5$ in comparison 
to $\eta=0.714$ ($\eta^{-3}=2.75$) as well as $\eta=0.909$ ($\eta^{-3}=1.331$), for which it is hardly visible anymore.

While the inner and outer cylinder boundary layers extend into the bulk equally for pure inner cylinder rotation (cf. \citep{ost14c}), 
the \emph{jump} of $\omega$ in the boundary layers is much larger in the inner cylinder 
as compared to the outer cylinder due to the different
slopes and equal extents. Therefore, the plumes are highly asymmetric, and smaller drivings $Ta$ break 
up the ``plume conveyor belts'', which form the large-scale structures seen in the time-averaged azimuthal velocity. 
On top of this plume asymmetry, originating from the boundary layers, a larger curvature
has an effect on the bulk. The underlying $\bar\omega(r)$ profile is less flat, and thus
the drop in angular velocity inside the bulk is the larger the smaller the value of $\eta$ is. 

\begin{figure}
 \begin{center}
  \subfloat{\label{fig:ommeEta05  }\includegraphics[height=0.40\textwidth,trim = 25mm 0mm 40mm 0mm, clip]{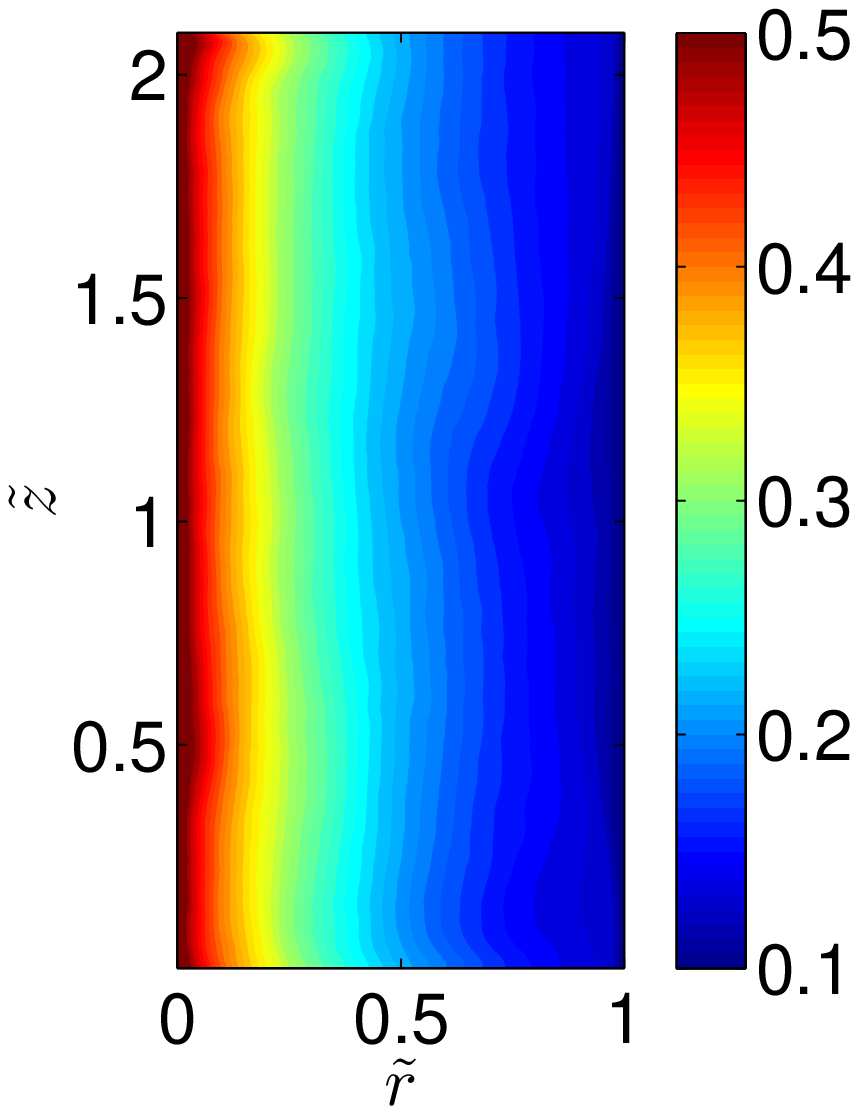}}
  \subfloat{\label{fig:ommeEta0714}\includegraphics[height=0.40\textwidth,trim = 25mm 0mm 40mm 0mm, clip]{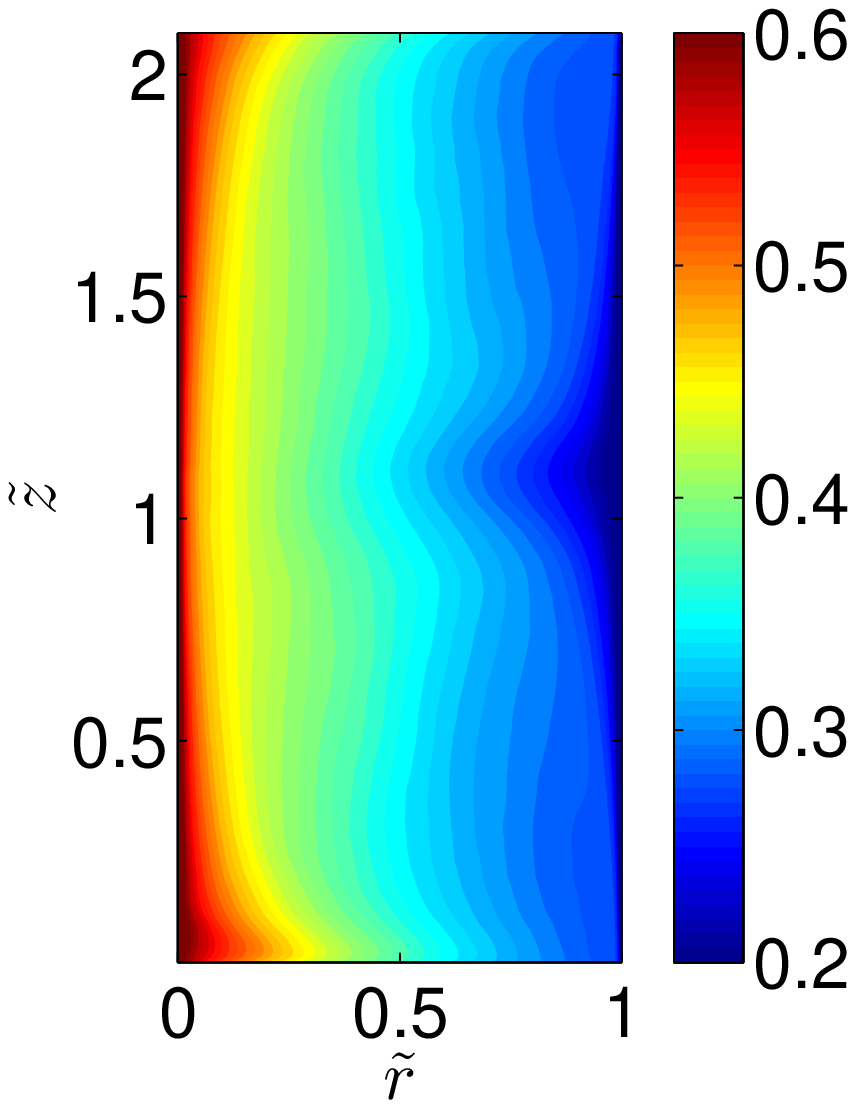}}
  \subfloat{\label{fig:ommeEta0909}\includegraphics[height=0.40\textwidth,trim = 25mm 0mm 40mm 0mm, clip]{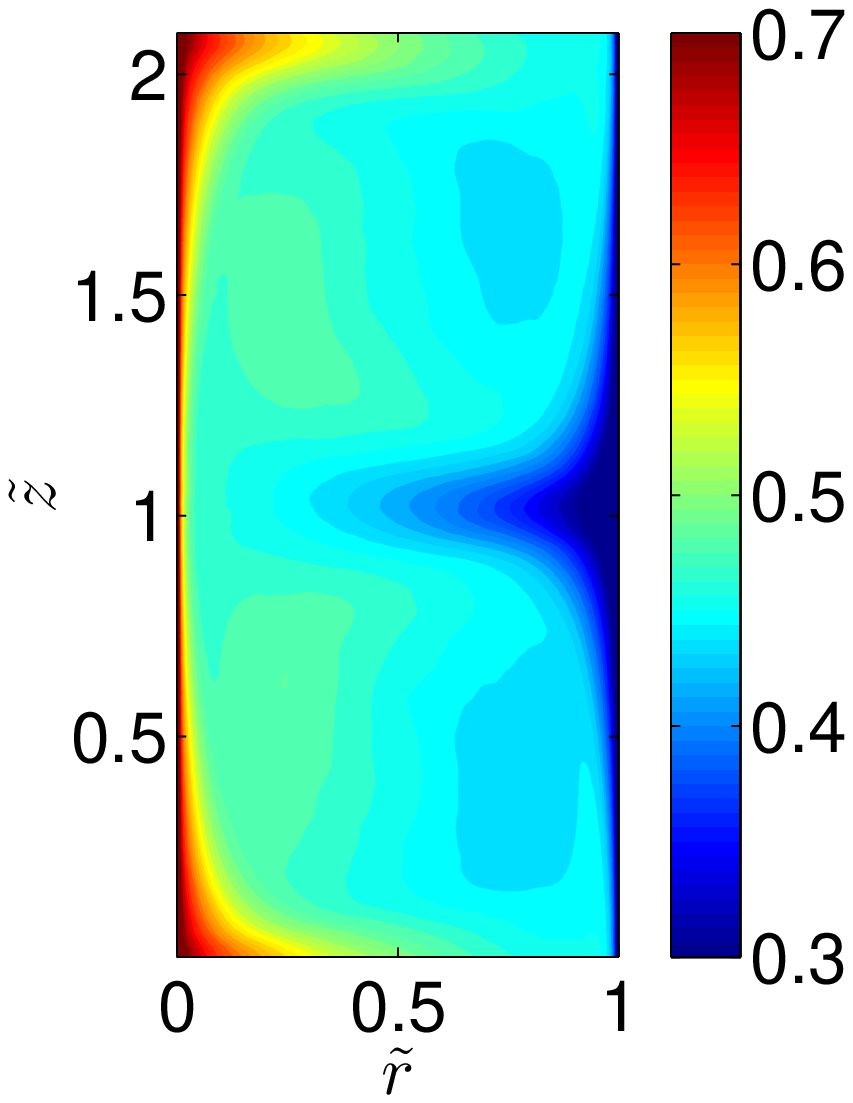}}
  \caption{Contour plots of the azimuthally- and time averaged angular velocity field $\bar{\omega}$ for $Ta = 10^{10}$ and $\usro=0$
and three values of $\eta$: $\eta=0.5$ (left), $\eta=0.714$ (middle) and $\eta=0.909$ (right). The colour scale has 
been shifted in order to account for the different bulk angular velocities at different $\eta$. Almost no axial 
dependence can be noticed for $\eta=0.5$, while it is
still very marked for $\eta=0.909$ }
  \label{fig:ommeEtas}
 \end{center} 
\end{figure}

\begin{figure}
 \begin{center}
  \includegraphics[width=0.47\textwidth]{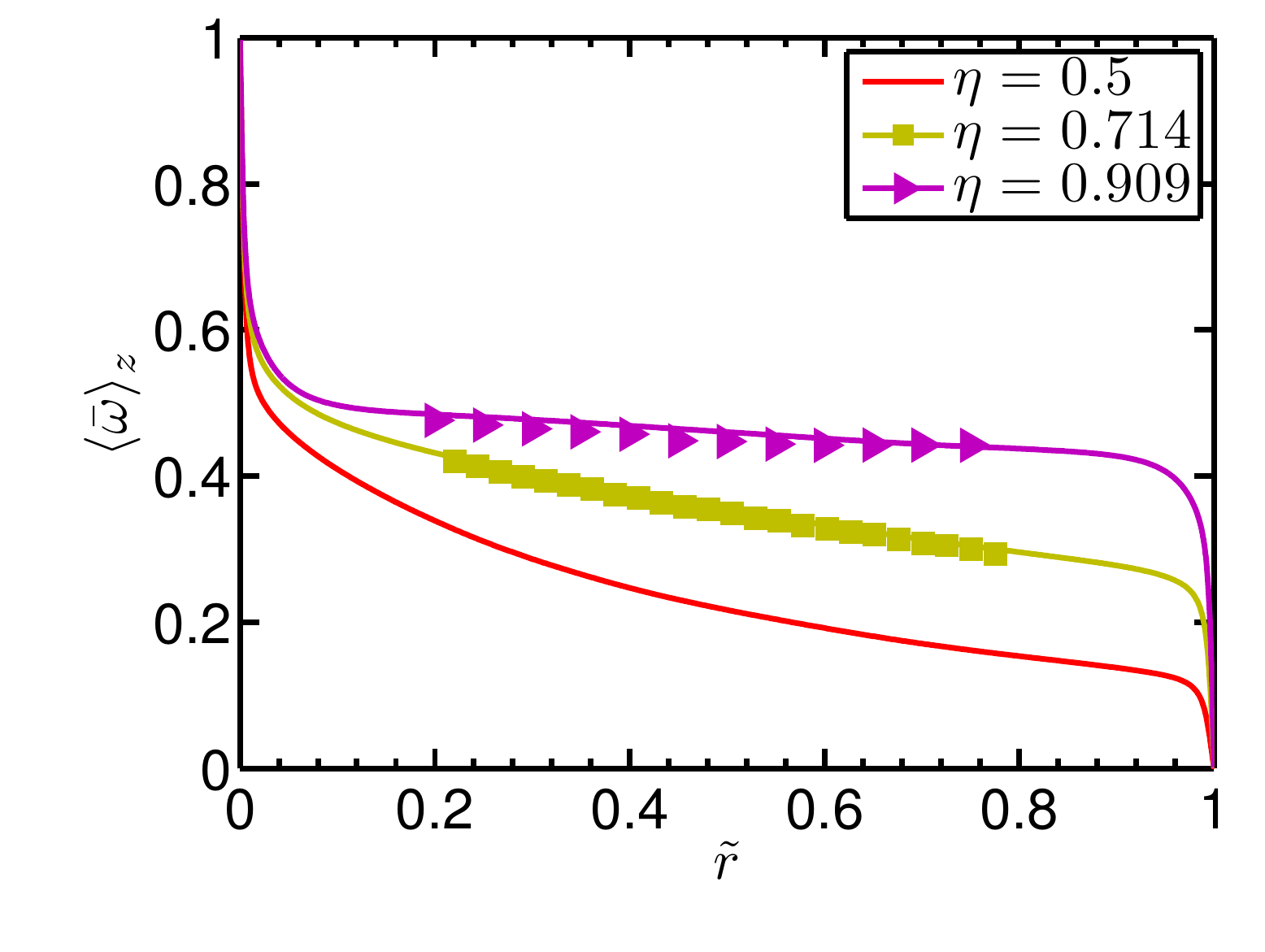}
  \includegraphics[width=0.47\textwidth]{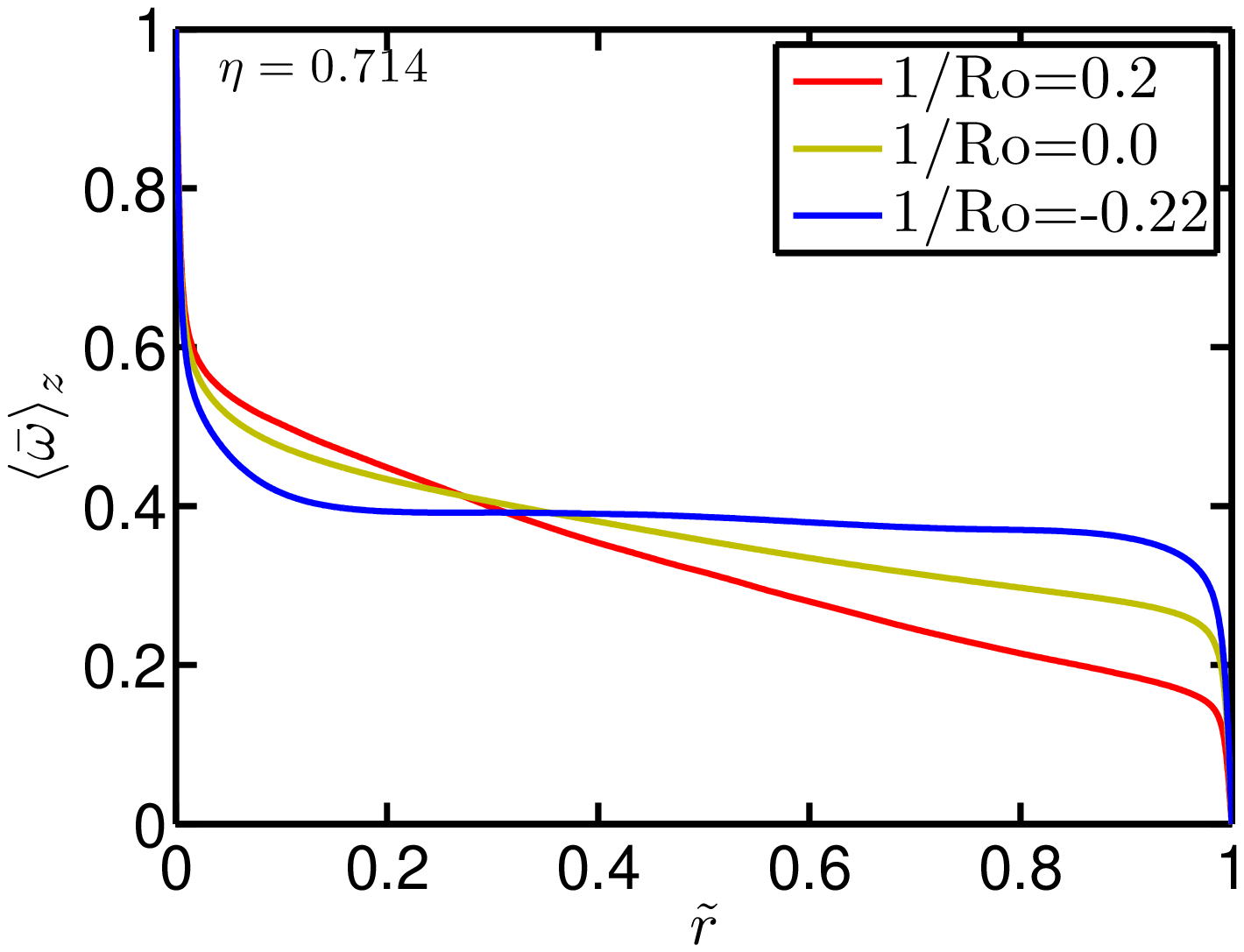}
  \caption{Left: Axially averaged angular velocity profiles $\langle\bar{\omega}\rangle_z$ for $\eta=0.5$, $\eta=0.714$, 
and $\eta=0.909$ at moderate driving $Ta=10^{10}$ and $\usro=0$. Solid lines are DNS data, while 
squares and triangles correspond to LDA data from experiments ($Ta=1.51\cdot10^{12}$ for $\eta=0.714$ and $Ta=1.1\cdot10^{11}$ 
for $\eta=0.909$)
\citep{ost14c}. A larger decrease of $\omega$ across the bulk can be seen for $\eta=0.5$. The angular velocity in the 
bulk also deviates more from $\omega=0.5$, the expected value in the limit case of $\eta\to 1$ (plane Couette flow). 
Right: Axially averaged angular velocity profiles $\langle\bar{\omega}\rangle_z$ for $\eta=0.714$, and three values 
of $\usro$ in the CWCR regime. The analogy
between the effects of $\eta$ and $\usro$ on $\bar{\omega}(\tilde{r})$ can be clearly seen.}
 \label{fig:OmegaZEtas}
 \end{center} 
\end{figure}

Both effects can be appreciated in figure \ref{fig:ommeEtas}, which shows contour plots of the azimuthally- and
time-averaged angular velocity $\bar{\omega}$ at $Ta=10^{10}$ for the three simulated values of $\eta$. This 
also explains the left panel of figure \ref{fig:OmegaZEtas}, where the now also axially 
averaged angular velocity $\langle \bar{\omega} \rangle_z$ 
is shown for the same three values of $\eta$. For comparison, the right panel of \ref{fig:OmegaZEtas} shows three profiles of
$\langle \bar{\omega} \rangle_z$ in the CWCR regime for $\eta=0.714$.

\begin{figure}
 \begin{center}
  \subfloat{\label{fig:TaDeltaUEtas}\includegraphics[width=0.49\textwidth]{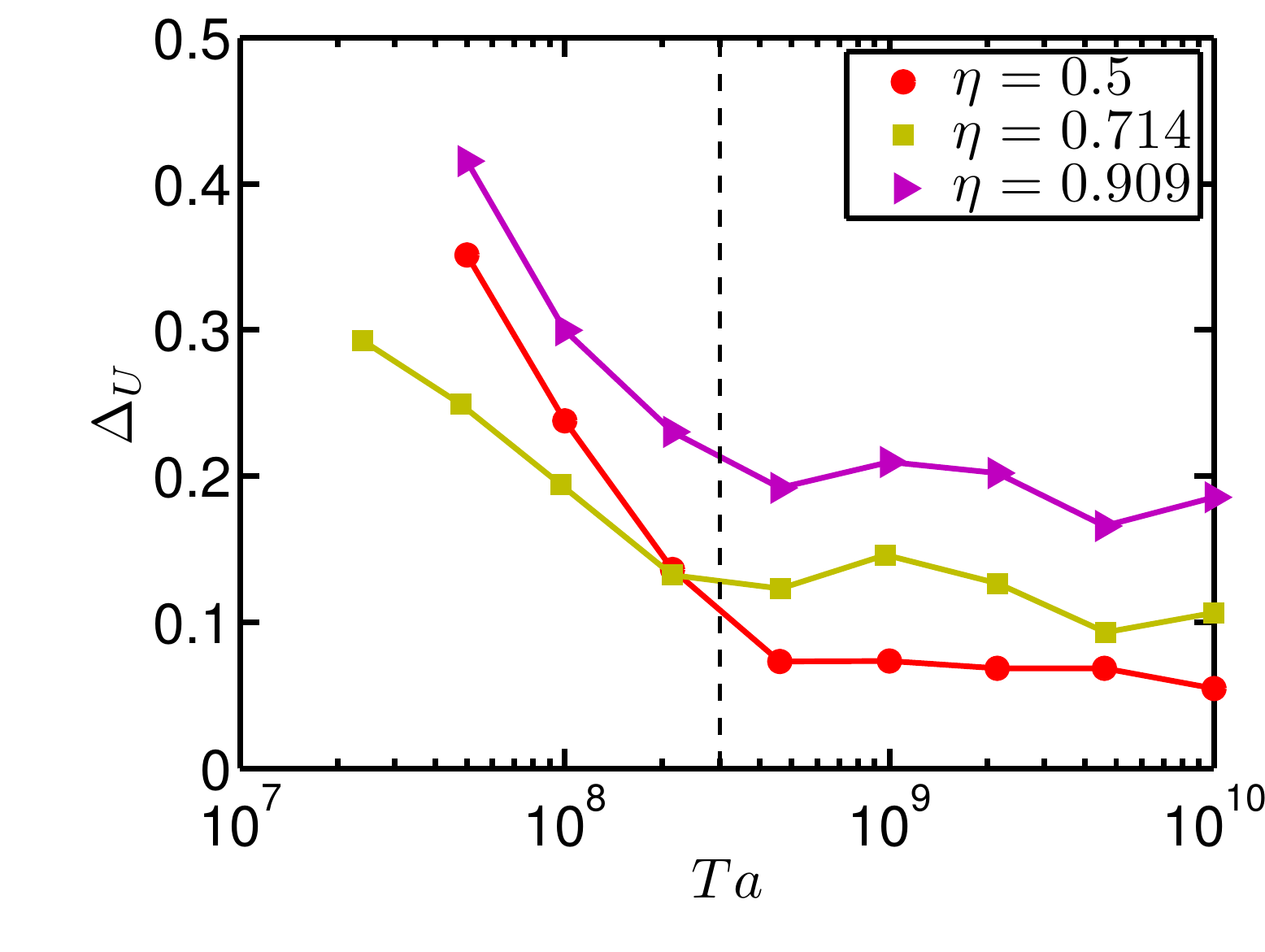}}                
  \caption{ The measure $\Delta_U$ for the axial velocity spread versus $Ta$ for the three values of $\eta$ simulated. 
A decrease in axial dependence can be
seen for all values of $\eta$ around $Ta\approx 10^8$, unlike what was seen for varying $\usro$, where the $Ta$ at which the 
decrease of axial dependence took place is $\usro$ dependent. However, the residual axial spread at 
the largest drivings increases with increasing $\eta$, 
as we would expect from the analogy between decreasing $\usro$ and increasing $\eta$.
  }
  \label{fig:TaDeltaUBLsEtas}
 \end{center} 
\end{figure}

\begin{figure}
 \begin{center}
  \includegraphics[width=0.49\textwidth]{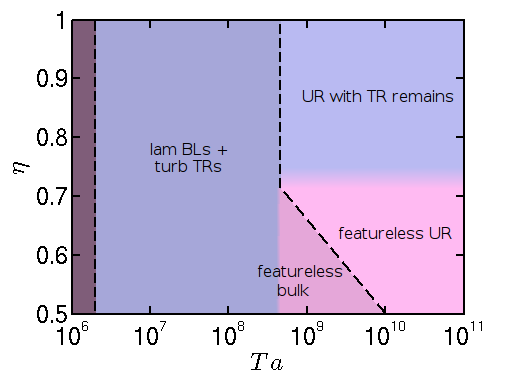}
  \caption{Transition between different regimes in the $(Ta,\eta)$ parameter space for pure inner
cylinder rotation $\usro=0$. The transition to the ultimate regime occurs at a higher $Ta$ for
smaller $\eta$ (wider gap), while the vanishing of the large scale structures occurs at around
the same $Ta$ for $0.5<\eta<0.714$. Remains of the Taylor rolls can only be seen for large $\eta$,
i.e. smaller gap. Abbreviations: boundary layer (BL), Taylor rolls (TR), ultimate regime (UR).  }
  \label{fig:phasespaceeta}
 \end{center} 
\end{figure}

The analogy between the effect of $\eta$ and the effect of $\usro$ on $\omega(\tilde{r})$ is also demonstrated in 
figure \ref{fig:OmegaZEtas}. 
The rolls are weak for $\eta=0.5$, as they are weak for co-rotating cylinders, and the rolls are strongest 
for $\eta=0.909$ and for $\usro\approx\usro_{opt}$. This also explains why, for large 
enough $Ta$, $\nom$ is highest at a given $Ta$ for the largest $\eta$.
However, the analogy is not perfect. For pure inner cylinder rotation, i.e., for $Ro^{-1} = 0$ the 
wide variety of flow states seen 
in \cite{and83} and \cite{and86} is greatly reduced. The system essentially goes from Taylor vortex flow
to modulated Taylor vortex flow to finally turbulent Taylor vortex flow. It does not undergo transitions
to different states (such as e.g. the ``wavelet'' state), and
thus the rolls do not vanish for the lower drivings at which this happens in co-rotating cylinders.
This can be seen in figure \ref{fig:TaDeltaUBLsEtas}, which shows the measure $\Delta_U$ for the axial velocity spread 
as function of $Ta$. With increased
driving, the rolls progressively lose importance until $Ta$ reaches a value of $Ta\approx 3\cdot10^8$. 
However, the effect of $\eta$, and thus of the cylinder wall curvature on the $\omega$ profiles 
can be clearly noticed in the \emph{residual} axial dependence and behaves as expected from the analogy.
The behaviour of the transition to the ultimate regime and associated sub--regimes
is summarized in figure \ref{fig:phasespaceeta}, which is analogous to figure \ref{fig:phasespacenew},
but now for the $(Ta,\eta)$ parameter space explored.

\begin{figure}
 \begin{center}
  \includegraphics[width=0.47\textwidth]{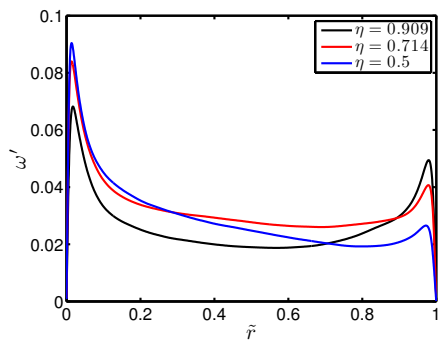}
  \caption{Root mean square (rms) profiles of the angular velocity fluctuations, $\omega'(\tilde{r})$, at $Ta=10^9$ and $\usro=0$
for the three $\eta$ values  simulated here. The boundary layer asymmetry causes the
fluctuations to be strongly reduced at the outer cylinder for $\eta=0.5$, as compared to those at the inner cylinder. }
  \label{fig:rmsEtas}
 \end{center} 
\end{figure}

Finally, one may ask the question of why the onset of the ultimate regime happens at a much higher $Ta$ for $\eta=0.5$ 
than for the two other values of $\eta$
studied. For $\eta=0.714$, the transition seems to set in for the same value of $Ta$ independently of $\usro$. 
A factor ten increase in shear in the boundary layers is required for 
the boundary layer instability to occur and the ultimate regime to set in.
Convex curvature is known to produce a stabilizing effect on boundary layers \citep{gor40,muc85}, 
and this will have a more significant effect on the inner cylinder for $\eta=0.5$ than for the larger $\eta$. 
On the other hand we might expect that the destabilizing
effect of concave curvature \citep{gor40b,hof85} would also play a role in accelerating the transition. Due to the  
boundary layer asymmetry however,
the outer boundary layer is much more ``quiet'', and has less fluctuations. This also delays the transition, and can 
be seen in figure \ref{fig:rmsEtas}, which shows the rms-fluctuations of the angular 
velocity $\omega^\prime=\langle\langle\omega^2\rangle_{t,\theta}-\bar{\omega}^2 \rangle^{1/2}_z$, 
for $Ta=10^9$ and the three values of $\eta$ simulated.
The levels of fluctuations at the outer cylinder are significantly reduced for $\eta=0.5$ when compared to the other 
values of $\eta$.
Finally, the large gradient of angular velocity sustained in the bulk will also reduce the shear in the outer cylinder, as the bulk angular velocity
is smaller for $\eta=0.5$. Thus, a combination of reduced fluctuations, stabilizing effect due to curvature at the inner cylinder, 
and reduced shear due to bulk angular velocity gradients is causing the delayed transition.

\section{Dependence on number and size of rolls}
\label{sec:lz}

Finally, we will quantify how the torque depends on the number and the size of the rolls, i.e. the 
vortical wavelength. The wavelength of a roll $\lambda_z$ is restricted to the values $\lambda=\Gamma/n$,
where $n$ is a strictly positive integer. For all simulations in this paper, $n=1$, and thus $\lambda=\Gamma$.
This is not necessarily always the case, $n$ is a response of the system,
and if $\Gamma$ is large enough, i.e. the system can accomodate more than one vortex pair, $n$ 
can take several values depending on how the final state of the system is reached.
\cite{bra13} showed
that for $\eta=0.714$, the ``optimal'' vortex wavelength, i.e. the vortex wavelength $\lambda_z$ which corresponds
to a maximum $\nom$, increased when comparing $\nom(\lambda_z)$ for two Taylor numbers, one in the Taylor vortex regime
and another in the turbulent Taylor vortex regime. For the higher $Ta$, the dependence of $\nom$ on $\lambda_z$ was quite weak.
\cite{mar14} showed that for $\eta=0.909$, different branches in the $\nom(Ta)$ relationship, associated to distinct
vortical states cross around $Re_i=1.3\cdot 10^4$. This corresponds to a driving of $Ta=1.8\cdot10^8$,
around the value at which the transition to the ultimate regime occurs for $\eta=0.909$.
The large-scale circulation could still be seen to play a role in determining the system response
after the transition to the ultimate regime. Furthermore, large scale patterns were observed in \cite{ost14} when looking 
at the $\langle \bar{\omega}\bar{u}_r\rangle$ correlation
at $Ta\sim 10^{10}$, even though they are absent when looking only at $\bar{\omega}$.

\begin{figure}
 \begin{center}
  \includegraphics[width=0.95\textwidth]{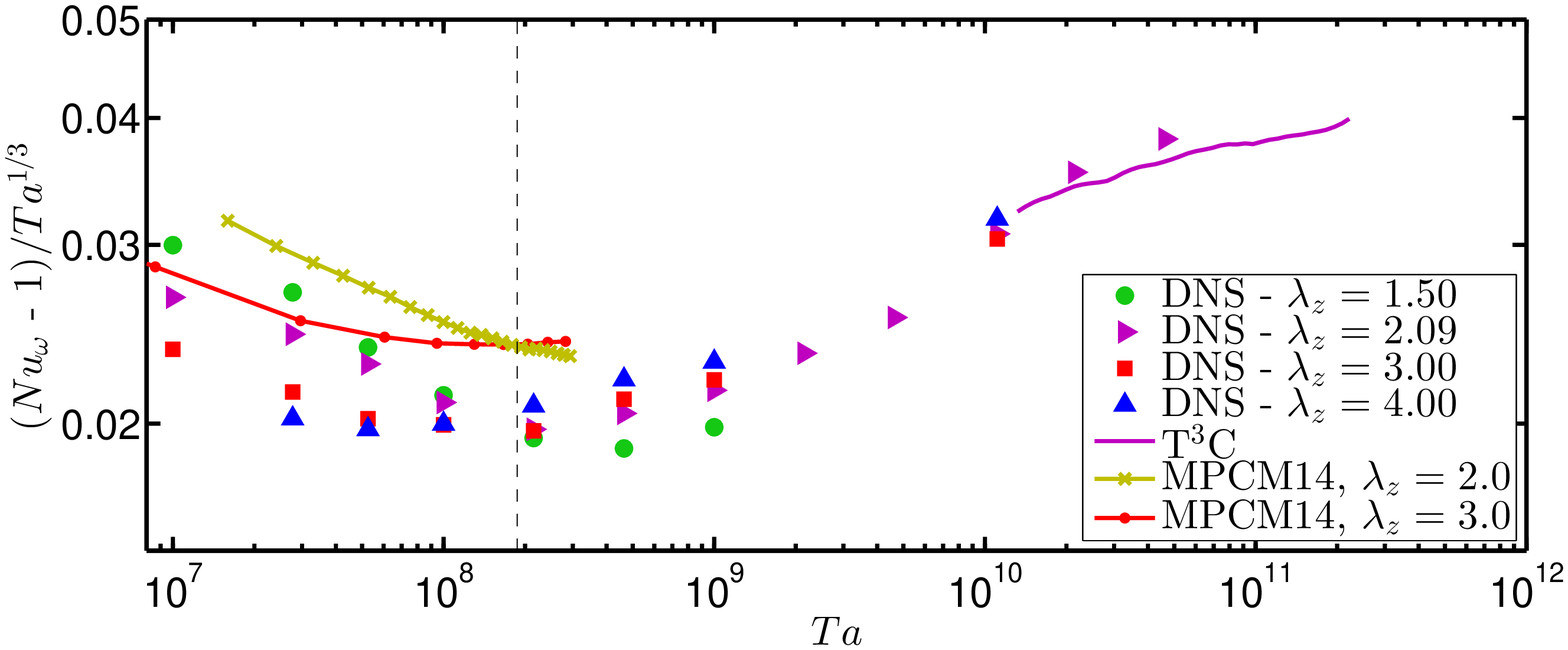}
  \caption{ Compensated torque $\nom$ versus driving strength $Ta$ for $\eta=0.909$ and the three different vortical
wavelengths. Experimental data from the $T^3C$ apparatus ($\Gamma=46.35$, number of rolls not
determined, cf. \cite{ost14c}) and from \cite{mar14}
(denoted MPCM14, $\lambda_z=2$ corresponds to 30 rolls and $\lambda_z=3$ corresponds to 18 rolls) are also plotted. 
Axial boundary conditions are different in experiments and DNSs. Experiments have end-plates, while
DNS are axially periodic and thus end effects are absent.
Both in experiment and in numerics, different branches associated to 
different states cross at $Ta\approx 2\cdot10^8$, shown as a vertical dashed line in the graph. This value 
of $Ta$ corresponds to the transition to the ultimate regime for radius ratio $\eta = 0.909$.
  }
  \label{fig:TaNuNNGammadep}
 \end{center} 
\end{figure}

Figure \ref{fig:TaNuNNGammadep} shows the compensated torque $\nom$ as function of $Ta$ for the four values of the vortical
wavelength studied. Experimental data by \cite{mar14} and DNS data by \cite{ost14c} is also plotted. It is worth noting 
that experimental data will have some end-plate effects, even if the aspect ratio $\Gamma$ of the experiments is larger than $30$, 
while the DNSs have periodic axial boundary conditions. Even so, very similar behaviour can be seen. 
The transition to the asymptotic scaling laws of the ultimate regime seem to occur 
around the same value of $Ta$, but are less pronounced the smaller the vortical wavelength is. 

The change in behaviour of the $\nom(Ta)$ curves
can be associated to the change of behaviour of the wind-sheared regions in the ultimate
regime.
As seen in \cite{ost14}, plume ejection is supressed outside the ultimate regime in 
regions of the flow, the so called ``wind-sheared'' regions due to the sweeping by the large scale rolls.
This reduction in plume ejection results in a reduced transport of angular velocity (torque). A similar reduction 
in the torque caused by a mean flow was also 
seen when forcing the flow with an axial pressure gradient by \cite{man09}.
Vortices with a smaller wavelength have smaller wind-sheared regions and thus result in a larger $\nom$,
if this suppression is taking place.
After the transition to the ultimate regime, the suppression ceases, and these regions 
become active ejectors of plumes, leading to increased transport. 

 The difference between  $\lambda_z=2.09$, $\lambda_z=3.0$ and the $\lambda_z=4.0$ is very small for $Ta=10^9$,
of the order of $5\%$, but for the $\lambda_z=1.5$ branch the difference is almost $15\%$. Only at $Ta=10^{10}$, when the distinction between wind-sheared and ejection regions is completely blurred away, and the whole inner cylinder can emit plumes (or hairpin vortices), $\nom(Ta)$ loses its $\Gamma$ dependence, within the error bars of the numerics. This sudden transition of wind-sheared regions to ejection regions causes the jump we see in the $\nom(Ta)$ curve at around $Ta=5\cdot 10^9$ for $\eta=0.909$.

Note that for the largest drivings axially periodic boundary conditions have been used, with only 
one vortex pair. This does not prevent the creation of two pairs of vortices with wavelength $\lambda_z=1.5$ 
by a breakup of one pair of vortices of $\lambda_z=3.0$ 
in a domain, which has $\Gamma=3.0$. And indeed this is seen to happen for the lower drivings both in DNS and
experiment. On the other hand, this axial periodicity affects the stability of one pair of vortices of wavelength $\lambda_z=1.5$
in a domain of $\Gamma=1.5$. Therefore, vortices with $\lambda_z=1.5$ might be an artifact
due to the numerical constraintment, and not be stable if a system with large $\Gamma$ at large $Ta$ is considered.
States with $\lambda_z<2$ are not reported in \cite{mar14}.

Even if we do not expect a quantitative agreement of the present DNS results
with those of \cite{bra13} and experimental data by \cite{hui14}, as we simulate a different $\eta$, 
the results reported in this section 
even do not agree qualitatively. \cite{bra13} see a maximum in torque for $\lambda_z=1.93$ 
in the turbulent Taylor vortex regime ($Ta\sim10^7$), 
while in the present simulations for $\eta=0.909$ at the same $Ta$, 
this maximum is clearly at $\lambda_z=1.5$, and not near $\lambda_z=2.09$. 
In the experiments of \cite{mar14}, states with $\lambda_z$ smaller than one are not reported, and a direct comparison cannot be made. 

We also note that the relationship between larger vortices and larger torque in the ultimate
regime is the inverse of what was recently reported by \cite{hui14}. 
\cite{hui14} found multiple states, with different $\lambda_z$ in highly turbulent TC flow.
For different states they found that the torque differs less than $5\%$, 
although they note that this might be due to the fact the torque is only measured on part of the inner cylinder, 
not on the entire inner cylinder. Furthermore their results are for $\usro\neq0$, for higher $Ta$, 
and for different $\eta$, as compared to the current research.

\section{Summary and conclusions}

Numerical simulations of turbulent Taylor-Couette flow in the range $10^4<Ta<4.6\cdot10^{10}$ were performed 
to explore the transition of TC flow to the (fully turbulent) ultimate 
regime. The four dimensions of the parameter space were explored, including the dependence of the transition on the 
radius ratio $\eta$, the vortex wavelength $\lambda_z$ and Coriolis force $Ro^{-1}$ or rotation ratio $\mu$.

First, the effect of the outer cylinder rotation, in the equations of motion in the frame
co-rotating with the outer cylinder, present as a Coriolis force, was analyzed 
for $\eta=0.714$. Depending on the value of $\usro$ two regimes were identified, (i) the co-rotating and 
weakly counter-rotating cylinder regime (CWCR) and (ii) the strongly counter-rotating cylinder regime (SCR),
both with their respective sub--regime. Our findings of that chapter culminate in the phase diagram fig. \ref{fig:phasespacenew},
in the $(Ta,\usro)$ regime (fig. \ref{fig:phasespacenew}a) and in the $(Re_i,Re_o)$ regime (fig. \ref{fig:phasespacenew}b \& c).
The transition to the ultimate regime could be observed for all values of $\usro$ around $Ta\sim 3\cdot10^8$.
However, for these two regimes a rather different behavior in the scaling laws $Nu_{\omega} (Ta)$ was found before 
the transition. We also found very different flow structures in the respective ultimate regimes in accordance with 
the description by \cite{bra13b}.
An explanation why the Coriolis force, proportional to $\usro$ stabilizes the large-scale structures was illustrated; the large-scale structures were
found to \emph{not} vanish at the transition to the ultimate regime for $\usro=-0.22\approx\usro_{opt}$, unlike what was seen
in \cite{ost14} for resting outer cylinder.

After this, the transition was analyzed for various gap widths, namely for $\eta=0.5$, $0.714$, and $0.909$ without 
Coriolis forces, i.e., for $\usro=0$. The transition was found to occur at about the same $Ta$ for $\eta=0.714$ and 
$0.909$. However, the transition was considerably delayed to $Ta\approx10^{10}$
for $\eta=0.5$, due to the combined effects of stabilizing curvature of the inner cylinder, 
and the reduced shear as well as smaller fluctuations in the vicinity of the outer cylinder. An analogy between the 
effect of $\usro$ in the CWCR regime and the effect of $\eta$ on the large scale rolls was described: 
Decreasing $\eta$ was found to have the same effect as adding a positive $\usro$ --corresponding to co-rotating 
cylinders-- , while increasing $\eta$ behaved like (weakly) counter-rotating the outer cylinder. 

Finally, as the large-scale structures were found to be strongest 
for $\eta=0.909$, the effect of varying the vortical wavelength was analysed for this value of $\eta$. 
As in \cite{mar14}, different branches of the $\nom(Ta)$ curve were found to cross around 
the transition to the ultimate regime. Before this transition,
the influence of the vortical wavelength (and thus of the aspect ratio) on $\nom$ was quite noticeable. After the ultimate 
range transition, this effect decreased drastically. The results of our DNS agree qualitatively
with those in the experiments by \cite{mar14} for $\eta=0.909$ even though the axial boundary conditions are different.
However, they are qualitatively different from those reported for $\eta=0.714$ by \cite{bra13} and by \cite{hui14}

In this work, the vortical wavelength by using periodic boundary conditions was fixed. Some of these states might not 
be accessible in experiment or might be a product of the periodic boundary conditions. 
Studying the coexistence of different states for large $\Gamma$, like
done in \cite{mar14} or \cite{hui14} with DNS requires a large amount of computational resources 
for high $Ta$. Switches between two and three vortex pairs
were seen at lower $Ta$ for $\eta=0.909$ \citep{ost14c}. Switching between states might also occur at high $Ta$, although
they are not captured in the DNS presented in this work. In the future,
additional DNS for $\eta=0.909$ with large $\Gamma$ at high $Ta$ should be run to improve the understanding 
of the switching between different states. 

Our ambition also is to further understand why the transition is delayed at $\eta=0.5$, 
but also the curvature effects on the $\omega$-profiles in the boundary layers along the ideas of \cite{gro14}. 
Curvature effects at $\eta=0.714$ and $\eta=0.909$ are too small to be
appreciated, and the flow for $\eta=0.5$ is still in the transition to the ``ultimate'' regime. Thus, higher 
$Ta$ simulations for $\eta=0.5$ will provide further understanding on how curvature makes the boundary layers of TC 
flow different from those of channel and pipe flow.

Acknowledgements: We would like to thank H. Brauckmann, J. Peixinho, M. Salewski, 
and C. Sun for various stimulating discussions during the years.
We would also like to thank the Dutch Supercomputing Consortium SurfSARA for technical support,
FOM, COST from the EU and ERC for financial support through an Advanced Grant. We acknowledge that these results
come from computational resources at the PRACE resource Curie, based in France at GENCI/CEA.

\bibliographystyle{jfm}

\bibliography{/home/rodosti/NetworkHomeCopy/literatur}

\appendix

\newpage

\textbf{APPENDIX: NUMERICAL DETAILS}

\begin{table}
  \begin{center} 
  \def~{\hphantom{0}}
  \begin{tabular}{|c|c|c|c|c|}
  \hline
  $Ta$   & $\usro$ & $\mu$ & $\nom$       & $N_\theta\times N_r \times N_z$ \\
  \hline
  $2.15\cdot10^8$    &  $0.20$ &  $0.2$ & $11.48$ & $256\times640\times512$  \\
  $2.15\cdot10^8$    & -$0.13$ & -$0.2$ & $13.43$ & $256\times640\times512$  \\
  $2.15\cdot10^8$    & -$0.22$ & -$0.4$ & $12.85$ & $256\times640\times512$  \\
  $2.15\cdot10^8$    & -$0.30$ & -$0.6$ & $11.13$ & $256\times640\times512$  \\
  $2.15\cdot10^8$    & -$0.40$ & -$1.0$ & $8.565$ & $256\times640\times512$  \\
  $4.64\cdot10^8$    &  $0.20$ &  $0.2$ & $14.21$ & $256\times640\times512$  \\
  $4.64\cdot10^8$    & -$0.13$ & -$0.2$ & $17.20$ & $256\times640\times512$  \\
  $4.64\cdot10^8$    & -$0.22$ & -$0.4$ & $17.77$ & $256\times640\times512$  \\
  $4.64\cdot10^8$    & -$0.30$ & -$0.6$ & $15.81$ & $256\times640\times512$  \\
  $4.64\cdot10^8$    & -$0.40$ & -$1.0$ & $11.36$ & $256\times640\times512$  \\
  $1.00\cdot10^9$    &  $0.20$ &  $0.2$ & $18.57$ & $256\times640\times512$  \\
  $1.00\cdot10^9$    & -$0.13$ & -$0.2$ & $23.10$ & $256\times640\times512$  \\
  $1.00\cdot10^9$    & -$0.22$ & -$0.4$ & $23.18$ & $256\times640\times512$  \\
  $1.00\cdot10^9$    & -$0.30$ & -$0.6$ & $19.85$ & $256\times640\times512$  \\
  $1.00\cdot10^9$    & -$0.40$ & -$1.0$ & $14.73$ & $256\times640\times512$  \\
  $2.15\cdot10^9$    &  $0.20$ &  $0.2$ & $24.96$ & $256\times640\times512$  \\
  $2.15\cdot10^9$    & -$0.13$ & -$0.2$ & $31.26$ & $256\times640\times512$  \\
  $2.15\cdot10^9$    & -$0.22$ & -$0.4$ & $31.41$ & $256\times640\times512$  \\
  $2.15\cdot10^9$    & -$0.30$ & -$0.6$ & $27.46$ & $256\times640\times512$  \\
  $2.15\cdot10^9$    & -$0.40$ & -$1.0$ & $20.15$ & $256\times640\times512$  \\
  $4.64\cdot10^9$    &  $0.20$ &  $0.2$ & $32.51$ & $384\times640\times768$  \\
  $4.64\cdot10^9$    & -$0.13$ & -$0.2$ & $41.44$ & $384\times640\times768$  \\
  $4.64\cdot10^9$    & -$0.22$ & -$0.4$ & $41.13$ & $384\times640\times768$  \\
  $4.64\cdot10^9$    & -$0.30$ & -$0.6$ & $36.39$ & $384\times640\times768$  \\
  $4.64\cdot10^9$    & -$0.40$ & -$1.0$ & $26.01$ & $384\times640\times768$  \\
  $1.00\cdot10^{10}$ &  $0.20$ &  $0.2$ & $41.01$ & $512\times800\times1024$ \\
  $1.00\cdot10^{10}$ & -$0.13$ & -$0.2$ & $57.50$ & $512\times800\times1024$ \\
  $1.00\cdot10^{10}$ & -$0.22$ & -$0.4$ & $58.61$ & $512\times800\times1024$ \\
  $1.00\cdot10^{10}$ & -$0.30$ & -$0.6$ & $49.98$ & $512\times800\times1024$ \\
  $1.00\cdot10^{10}$ & -$0.40$ & -$1.0$ & $34.42$ & $512\times800\times1024$ \\
  $2.15\cdot10^{10}$ & $0$     & $0$    & $66.57$ & $768\times1024\times1536$ \\
  $4.64\cdot10^{10}$ & $0$     & $0$    & $94.77$ & $768\times1200\times2048$ \\          
 \hline
 \end{tabular}
 \caption{This table presents a summary of the numerical results for $\eta=0.714$ which are new to this manuscript.
For the other data points see \cite{ost14}. The first column shows the driving, $Ta$. The second and third column show
the outer cylinder rotation as either a Coriolis force $\usro$ or a rotation frequency ratio $\mu=\omega_o/\omega_i$. 
The fourth column shows the non-dimensionalized torque, $\nom$. The fifth column shows the amount of grid points 
used in azimuthal ($N_\theta$), radial ($N_r$) and axial direction ($N_z$). All these simulations use a rotational
symmetry order six in the azimuthal direction, and are for $\Gamma=2.09$}
 \label{tbl:final}
\end{center}
\end{table}

\begin{table}
  \begin{center} 
  \def~{\hphantom{0}}
  \begin{tabular}{|c|c|c|c|c|c|}
  \hline
  $Ta$        & $\eta$   & $\Gamma$ & $\lambda_z$  & $\nom$ & $N_\theta\times N_r \times N_z$ \\
  \hline
  $2.15\cdot10^8$    & $0.5$    & $2.09$ & $2.09$ & $9.33$ & $384\times512\times768$ \\
  $4.64\cdot10^8$    & $0.5$    & $2.09$ & $2.09$ & $11.9$ & $384\times701\times768$ \\
  $1.00\cdot10^9$    & $0.5$    & $2.09$ & $2.09$ & $14.9$ & $512\times768\times768$ \\
  $2.15\cdot10^9$    & $0.5$    & $2.09$ & $2.09$ & $18.8$ & $768\times768\times1024$ \\
  $4.64\cdot10^9$    & $0.5$    & $2.09$ & $2.09$ & $24.1$ & $768\times768\times1024$ \\
  $1.00\cdot10^{10}$ & $0.5$    & $2.09$ & $2.09$ & $31.3$ & $1024\times1024\times1536$ \\
  $2.15\cdot10^{10}$ & $0.5$    & $2.09$ & $2.09$ & $40.9$ & $1024\times1024\times1536$ \\
  $4.64\cdot10^{10}$ & $0.5$    & $2.09$ & $2.09$ & $53.9$ & $1024\times1024\times2048$ \\
  $2.76\cdot10^7$    & $0.909$  & $2.09$ & $2.09$ & $12.8$ & $256\times512\times480$ \\
  $5.26\cdot10^7$    & $0.909$  & $2.09$ & $2.09$ & $16.8$ & $256\times512\times480$ \\
  $1.00\cdot10^9$    & $0.909$  & $2.09$ & $2.09$ & $22.6$ & $512\times768\times768$ \\
  $2.15\cdot10^9$    & $0.909$  & $2.09$ & $2.09$ & $31.3$ & $512\times768\times768$ \\
  $4.64\cdot10^9$    & $0.909$  & $2.09$ & $2.09$ & $43.6$ & $1024\times768\times768$ \\
  $1.00\cdot10^{10}$ & $0.909$  & $2.09$ & $2.09$ & $67.2$ & $1024\times1024\times1024$ \\
  $2.15\cdot10^{10}$ & $0.909$  & $2.09$ & $2.09$ & $99.3$ & $1536\times1536\times1024$ \\
  $4.64\cdot10^{10}$ & $0.909$  & $2.09$ & $2.09$ & $138$ & $2048\times1536\times1024$ \\
  $1.00\cdot10^6$    & $0.909$  & $1.50$ & $1.50$ & $4.31$ & $256\times512\times480$ \\
  $1.00\cdot10^7$    & $0.909$  & $1.50$ & $1.50$ & $7.46$ & $256\times512\times480$ \\
  $2.76\cdot10^7$    & $0.909$  & $1.50$ & $1.50$ & $9.15$ & $256\times512\times480$ \\
  $5.26\cdot10^7$    & $0.909$  & $1.50$ & $1.50$ & $9.91$ & $256\times512\times480$ \\
  $1.00\cdot10^8$    & $0.909$  & $1.50$ & $1.50$ & $10.9$ & $256\times512\times480$ \\
  $2.15\cdot10^8$    & $0.909$  & $1.50$ & $1.50$ & $12.6$ & $256\times512\times480$ \\
  $4.64\cdot10^8$    & $0.909$  & $1.50$ & $1.50$ & $15.6$ & $256\times512\times480$ \\
  $1.00\cdot10^9$    & $0.909$  & $1.50$ & $1.50$ & $20.8$ & $512\times512\times480$ \\
  $1.00\cdot10^6$    & $0.909$  & $3.00$ & $3.00$ & $3.60$ & $256\times512\times480$ \\
  $1.00\cdot10^7$    & $0.909$  & $3.00$ & $3.00$ & $6.10$ & $256\times512\times480$ \\
  $2.76\cdot10^7$    & $0.909$  & $3.00$ & $3.00$ & $7.50$ & $256\times512\times480$ \\
  $5.26\cdot10^7$    & $0.909$  & $3.00$ & $3.00$ & $8.58$ & $256\times512\times480$ \\
  $1.00\cdot10^8$    & $0.909$  & $3.00$ & $3.00$ & $10.3$ & $256\times512\times480$ \\
  $2.15\cdot10^8$    & $0.909$  & $3.00$ & $3.00$ & $12.8$ & $256\times512\times480$ \\
  $4.64\cdot10^8$    & $0.909$  & $3.00$ & $3.00$ & $17.4$ & $256\times512\times480$ \\
  $1.00\cdot10^9$    & $0.909$  & $3.00$ & $3.00$ & $23.1$ & $512\times512\times720$ \\
  $1.11\cdot10^{10}$ & $0.909$  & $3.00$ & $3.00$ & $68.9$ & $1024\times1024\times3072$ \\
  $2.76\cdot10^7$    & $0.909$  & $4.00$ & $4.00$ & $7.12$ & $256\times512\times480$ \\
  $5.26\cdot10^7$    & $0.909$  & $4.00$ & $4.00$ & $8.40$ & $256\times512\times480$ \\
  $1.00\cdot10^8$    & $0.909$  & $4.00$ & $4.00$ & $10.3$ & $256\times512\times480$ \\
  $2.15\cdot10^8$    & $0.909$  & $4.00$ & $4.00$ & $13.5$ & $256\times512\times480$ \\
  $4.64\cdot10^8$    & $0.909$  & $4.00$ & $4.00$ & $18.1$ & $256\times512\times480$ \\
  $1.00\cdot10^9$    & $0.909$  & $4.00$ & $4.00$ & $24.0$ & $512\times512\times720$ \\
  $1.11\cdot10^{10}$ & $0.909$  & $4.00$ & $4.00$ & $69.8$ & $2048\times1024\times4096$ \\
 \hline
 \end{tabular}
 \caption{This table presents a summary of the numerical results for the various geometries at $\usro=0$, i. e., 
for resting outer cylinder, which are new to this manuscript.
For the other data points see \cite{ost13} and \cite{ost14c}. The first column shows the driving, $Ta$. The second and third column show
the radius ratio $\eta$ and the aspect ratio $\Gamma$. The fourth column shows the vortical wavelength $\lambda_z$.
The fifth column shows the non-dimensionalized torque, $\nom$.
The sixth column shows the amount of grid points used in azimuthal ($N_\theta$), radial ($N_r$) and axial direction ($N_z$). We note that the $Ta=1.11\cdot10^{10}$, $\Gamma=4$, $\eta=0.909$ was done using $n_{sym}=10$.}
 \label{tbl:final2}
\end{center}
\end{table}

\end{document}